\documentclass[prd,aps,amssymb,amsmath,amsfonts,11pt]{revtex4}
\usepackage{latexsym}
\usepackage[latin1]{inputenc}
\usepackage[dvips]{graphicx}
\usepackage{color}

\newcommand{\ba}{\begin{eqnarray}}
\newcommand{\ea}{\end{eqnarray}}

\newcommand{\be}{\begin{equation}}
\newcommand{\ee}{\end{equation}}
\newcommand{\pa}{\partial}

\newcommand{\HQSS}{{\rm HQSS}}
\newcommand{\SU}{\mbox{SU}}
\usepackage{dcolumn}

\begin{document}
                    
\title{Transport coefficients of heavy baryons}
\author{Laura Tolos}
\affiliation{Instituto de Ciencias del Espacio (IEEC/CSIC), Campus Universitat Aut\`onoma de Barcelona, Carrer de Can Magrans, s/n, 08193 Cerdanyola del Vall\`es,
Spain\\
Frankfurt Institute for Advanced Studies. Johann Wolfgang Goethe University, Ruth-Moufang-Str. 1,
60438 Frankfurt am Main, Germany }
\author{Juan M. Torres-Rincon}
\affiliation{Subatech, UMR 6457, IN2P3/CNRS, Universit\'e de Nantes, \'Ecole de Mines de Nantes, 4 rue Alfred Kastler 44307,
Nantes, France}
\author{Santosh K. Das}
\affiliation{Department of Physics and Astronomy, University of Catania,
Via S. Sofia 64, 1-95125 Catania, Italy \\
Laboratori Nazionali del Sud, INFN-LNS, Via S. Sofia 62, I-95123 Catania, Italy }

\begin{abstract}
We compute the transport coefficients (drag and momentum diffusion) of the low-lying heavy baryons 
$\Lambda_c$ and $\Lambda_b$ in a medium of light mesons formed at the later stages of high-energy heavy-ion collisions. We employ the
Fokker-Planck approach to obtain the transport coefficients from 
unitarized baryon-meson interactions based on effective field theories that respect chiral and heavy-quark symmetries.
We provide the transport coefficients as a function of temperature and heavy-baryon momentum, and analyze the
applicability of certain nonrelativistic estimates. Moreover we compare our outcome for the spatial diffusion
coefficient to the one coming from the solution of the Boltzmann-Uehling-Uhlenbeck transport equation and we find a very good agreement
between both calculations. The transport coefficients for $\Lambda_c$ and $\Lambda_b$ in a thermal bath will be
used in a subsequent publication as input in a Langevin evolution code for the generation and propagation of heavy
particles in heavy-ion collisions at LHC and RHIC energies.
\end{abstract}

\maketitle
\newpage
\tableofcontents

\section{Introduction}

The energy dissipation of heavy quarks is considered as one of the most 
promising probes for the characterization of the plasma of quarks and gluons (QGP) formed in the 
early stages of heavy-ion collisions~\cite{Rapp:2009my}. The two main 
processes through which a heavy quark loses energy are elastic collisions~\cite{Svetitsky:1987gq,Braaten:1991we} and 
the bremsstrahlung or radiative loss~\cite{Mustafa:1997pm,Dokshitzer:2001zm} due to the interaction of the heavy quarks 
with the light quarks, light antiquarks and gluons in the QGP. However, to make the characterization of 
QGP reliable, the role of the hadronic phase should be also taken into consideration. 

Heavy hadrons, created after hadronization, suffer from collisions with light mesons
(such as $\pi$, $K$, $\bar K$ and $\eta$ mesons), rearranging their momentum until freeze-out and, thus, creating
a nuclear modification ratio ($R_{AA}$)  different from one as well as a finite elliptic flow ($v_2$). Recent results for $D$ mesons
(also for $D^{*}$ mesons) have been published by the ALICE Collaboration focusing on $R_{AA}$ \cite{Adam:2015sza} and $v_2$ \cite{Abelev:2014ipa}. However, measurements are still in an early stage, with several results in $p+p$ collisions 
and some recent results from CMS Collaboration in $p+Pb$ collisions at $\sqrt{s_{NN}}=5.02$ TeV~\cite{Khachatryan:2015uja}.

Heavy baryons, such as $\Lambda_c$ and $\Lambda_b$ baryons, could be experimentally accessible in the near future.
Initial analysis have been done for the $\Lambda_c$ baryon in $p+p$ collisions by the LHCb
Collaboration~\cite{Aaij:2013mga}. Although this state is very difficult to reconstruct due to its three-body decay,
there exist plans to study it within the Run 3 of LHC~\cite{Andronic:2015wma}. In particular, the ALICE
Collaboration will study the $\Lambda_c/D$ ratio, as well as $v_2$ and $R_{AA}$ after the upgrade of the Inner Tracking
System~\cite{Abelevetal:2014dna}, allowing for an improvement in the detection of heavy hadrons. The bottom counterpart 
$\Lambda_b$ is, however, more elusive, but its future reconstruction in Run 3 by ALICE has been 
considered~\cite{Abelevetal:2014dna,Andronic:2015wma}. In addition, this state has been recently measured by LHCb 
Collaboration in the context of a possible pentaquark-mediated decay~\cite{Aaij:2015tga}.

The physical observables, such as particle ratios, $R_{AA}$ or $v_2$, are strongly correlated to the behavior of 
the transport properties of heavy hadrons. The transport properties depend crucially on the interaction of the heavy particles with
the surrounding medium. From the theoretical point of view, these interactions have to be rigorously derived from effective field
theories (EFTs). EFTs are based on perturbative expansions exploiting some approximate symmetries of the system (e.g. chiral symmetry
and/or heavy-quark symmetries). In some cases, one can obtain model-independent results that correctly describe the low-energy
limit of QCD for the hadronic degrees of freedom. However, in practice, the validity of these EFTs for describing the
hadronic interactions is usually limited for very low energies/temperatures. The presence of resonant states breaks the
perturbative expansion and, as a consequence, unitarization of the effective theory is required. Unitarized effective
theories provide scattering amplitudes that respect the exact unitarity condition and can be then used for a reliable
determination of the transport properties of heavy hadrons.

Following the initial works of Refs.~\cite{Laine:2011is,He:2011yi,Ghosh:2011bw,Abreu:2011ic,Das:2011vba}, we have exploited chiral 
and heavy-quark symmetries to obtain the effective interaction
of heavy mesons, such as $D$~\cite{Tolos:2013kva} and $\bar B$~\cite{Torres-Rincon:2014ffa} mesons, with light mesons and
baryons. With these interactions, we have obtained the heavy-meson transport coefficients as a function of
temperature and baryochemical potential of the hadronic bath by means of solving the Fokker-Planck equation. In the present
work we move forward and calculate the transport coefficients of the low-lying heavy baryons ($\Lambda_c$ and $\Lambda_b$) using
a similar framework to account for the interaction of these states with light mesons ($\pi,K,\bar{K},\eta$). These light
mesons are expected to be the main components of the hadronic medium produced in high-energy heavy-ion collisions at
LHC or RHIC energies. We, moreover, investigate the applicability of certain estimates of the transport coefficients
based on the Einstein relation in the nonrelativistic regime. We finally compare the spatial diffusion coefficient
within the Fokker-Planck formalism to the one resulting from the solution of the Boltzmann-Uehling-Uhlenbeck
equation.

We leave the phenomenological implications of our findings for a subsequent paper in which we will accommodate the effective 
interactions of heavy baryons together with the initial dynamics of $c$ and $b$ quarks. We will extract several
observables potentially measurable at LHC energies, like the $R_{AA}$, $v_2$,
and baryon-to-meson ratios. In particular, we will provide predictions for $\Lambda_c$ baryons highlighting the impact 
of the hadronic medium interactions, which
are expected to be measured by the ALICE Collaboration. In addition, we will also
present predictions for the $\Lambda_b$ baryon, in spite of its more difficult experimental reconstruction.

The paper is organized as follows. In Sec.~\ref{sec:EFT} we describe the meson-baryon effective
interactions used to compute the scattering amplitudes and cross sections of $\Lambda_c$ and $\Lambda_b$ baryons 
in a light-meson bath. In addition, we review the interactions of $D$ and $\bar B$ mesons in the same medium.  
In Sec.~\ref{sec:transport} we present
the theory of the transport coefficients using the Fokker-Planck approach. Moreover, we show our results for the 
transport coefficients for $\Lambda_c$ and $\Lambda_b$, and investigate the applicability of certain nonrelativistic 
estimates of these coefficients.
We, furthermore, compare our result for the spatial diffusion coefficient with the one resulting from a direct
solution of the Boltzmann equation, as described in Appendix \ref{app:BUU}. 
Finally, in Sec.~\ref{sec:conclusions}, we present our conclusions.

\section{Effective hadronic interaction and unitarization~\label{sec:EFT}}

In this section we review the details of the interaction of $\Lambda_c$ and $\Lambda_b$ baryons with light mesons, such
as $\pi,K,\bar{K},\eta$. We also recall the details for the interaction of $D$ and $\bar B$ mesons for the same set
of light mesons. In general, any elastic scattering of hadrons is represented as
\be \label{eq:collision} H (p) \ l (q) \rightarrow H (p-k) \ l (q+k) \ , \ee
where $H$ and $l$ will denote the heavy and light states, respectively, with the corresponding four-momenta, for instance
$p^\mu = (p^0,{\bf p})$. 

The hadronic interactions are based on effective field theories where the relevant 
symmetries of the system are exploited. In the present case, heavy-quark and chiral symmetries are 
the fundamental guiding principles we use to construct the effective Lagrangian. As we are interested in realistic  
cross sections at moderate energies---where the presence of resonant states should be taken into account---we 
unitarize the scattering amplitudes so that exact unitarity  is imposed.
We generate the transition probabilities and cross sections that are needed for the microscopical 
calculation of the transport coefficients.

\subsection{$D$ and $\bar B$ mesons interacting with light mesons~\label{sec:meson}}

  In this section we briefly review the interaction between heavy mesons ($D,\bar{B}$) and light mesons
($\pi,K,\bar{K},\eta$), as we will use their transport coefficients to check some estimates.
In Ref.~\cite{Abreu:2011ic} we introduced the effective Lagrangian that fulfills chiral
symmetry constraints, and also respects heavy-quark symmetries when heavy mesons are present. The Lagrangian is
expanded up to next-to-leading order (NLO)  in the chiral expansion to account for the finite masses of the light mesons.
In addition, the heavy-quark mass expansion is always kept at leading order.  

The tree-level scattering amplitude for the process of Eq.~(\ref{eq:collision}) (but not restricted to elastic collisions) for total isospin $I$, spin $J$ and
strangeness $S$  is given by
\ba V^{IJS} (p,q,k) &=&  
\frac{C_0}{4f_\pi^2} [(2p -k) \cdot (2q+k)] + \frac{2C_1 h_1}{f_\pi^2} + \frac{2C_2}{f_\pi^2} h_3 (q \cdot q + q \cdot k)  \label{eq:potm} \\
 & + & \frac{2C_3}{f_\pi^2} h_5 [ 2(p \cdot q)^2 + 2 (p \cdot q) (p\cdot k) - 2 (p \cdot q) (q\cdot k) - (p \cdot k)(q\cdot k) - (p \cdot q) (k\cdot k)
] \ , \nonumber \ea
with $C_i$ the isospin coefficients of Ref.~\cite{Torres-Rincon:2014ffa} (which depend on $I,J,S$). The amplitude (also
known as potential) is  functionally independent of the heavy meson species ($D$ or $\bar{B}$) by virtue of the heavy-quark flavor symmetry (only broken
by the different $h_i$ coefficients). The pion decay constant is denoted as $f_\pi$, whereas
the $h_i$ are the so-called low-energy constants that only appear in the NLO terms. Three of the low-energy constants ($h_1,h_3$ and $h_5$) are free parameters and need to be
adjusted to experimental or lattice data~\cite{Abreu:2011ic,Abreu:2012et}. 

While in Ref.~\cite{Abreu:2011ic} we focused on the $D\pi$ scattering neglecting any other meson-meson channel, in a subsequent 
work for the bottom sector~\cite{Abreu:2012et} we considered the interaction of $B$ with all the
(pseudo)Goldstone bosons $\pi$, $K$, $\bar{K}$ and $\eta$ mesons. However, we neglected any possible mixing 
between meson-meson asymptotic states (an effect which is subleading in chiral counting). The inclusion of the coupled channel
$D\pi \leftrightarrow D\eta$ for the $D$ meson was considered in Ref.~\cite{Tolos:2013kva}, together with the
interaction of $D$ mesons with light baryons ($N$ and $\Delta$). Finally, in Ref.~\cite{Torres-Rincon:2014ffa},  we
accounted for all possible $\bar B$--light meson and $\bar B$--light baryon coupled channels, providing the most complete 
calculation of the scattering amplitudes.

 In the present work we follow the steps of Ref.~\cite{Torres-Rincon:2014ffa} i.e. we consider all possible meson-meson channels
for the $D$ meson scattering off $\pi$, $K$, $\bar K$ and $\eta$ in a consistent coupled-channel analysis.
We thus recalculate the low-energy constants to reproduce the
pole position of the $D_0(2400)$ resonance, and the mass splitting between $D$ and $D_s$. The
low-energy constants we use are $h_1=-0.45$, $h_3=5.5$, $h_5=-0.45$ GeV$^{-2}$. For the $\bar B$-light meson scattering,
we use the same results of Ref.~\cite{Torres-Rincon:2014ffa} with the parameters quoted there. Thus, both charm and
bottom sectors are computed with the same level of consistency. 

\subsection{$\Lambda_c$ and $\Lambda_b$ baryons interacting with light mesons}

The interaction of $\Lambda_c$ and $\Lambda_b$ scattering off $\pi$, $K$, $\bar K$ and $\eta$ mesons is obtained
 within a unitarized meson-baryon coupled-channel model that incorporates heavy-quark spin symmetry (HQSS)
~\cite{GarciaRecio:2008dp, Gamermann:2010zz, Romanets:2012hm,GarciaRecio:2012db,Garcia-Recio:2013gaa,Tolos:2013gta}.
This is a predictive model for four flavors including all basic hadrons (pseudoscalar and vector mesons, and $1/2^+$
 and $3/2^+$ baryons) which reduces to the Weinberg-Tomozawa (WT) interaction in the sector where Goldstone bosons
 are involved. This scheme has $\SU(6)\times \HQSS$ symmetry, i.e., spin-flavor symmetry in the light sector and HQSS in the heavy (charm/bottom) sector, and it is consistent with chiral symmetry in the light sector.

The extended WT meson-baryon interaction in the coupled meson-baryon basis reads
\begin{equation}
V_{ij}^{IJS} =
{D}_{ij}^{IJS}\,\frac{2\sqrt{s}-M_i-M_j}{4f_if_j} 
\sqrt{\frac{E_i+M_i}{2M_i}}\sqrt{\frac{E_j+M_j}{2M_j}}, 
\label{eq:potb}
\end{equation}
where $\sqrt{s}$ is the center-of-mass (C.M.) energy of the system; $E_i$ and $M_i$ are, respectively, the C.M. energy and mass of the baryon in the channel $i$; and $f_i$ is the decay constant of the meson in the $i$-channel.  The explicit breaking of the symmetries is achieved using the hadron masses and meson decay constants of Ref.~\cite{Romanets:2012hm,GarciaRecio:2012db}, while the ${D}^{IJS}_{ij}$ are given in Refs.~\cite{GarciaRecio:2008dp, Gamermann:2010zz, Romanets:2012hm,Garcia-Recio:2013gaa}. 

We are interested in analyzing the interaction of $\Lambda_c$ ($\Lambda_b$) with $\pi$, $K$, $\bar K$ and $\eta$. Thus, one needs to study the interaction in ($ I=1, J=1/2,S=0$) sector for $\Lambda_c(\Lambda_b) \pi$; ($I=1/2, J=1/2,S=1,-1$) for $\Lambda_c(\Lambda_b)K$ and $\Lambda_c(\Lambda_b) \bar K$, respectively, whereas for $\Lambda_c(\Lambda_b) \eta$ one analyzes the ($I=0, J=1/2,S=0$) sector.

\subsection{Unitarized effective interactions}
In order to obtain the invariant matrix elements that appear in the analysis of the transport coefficients, one has to
calculate the scattering amplitudes $T_{ij}$. We solve the on-shell 
Bethe-Salpeter equation (BSE) \cite{Oller:1997ti,Oller:2000fj}, using $V$ as kernel
\be
T_{ij}=[1 -  V G ]_{ik}^{-1} \ V_{kj} \ ,
\ee
where $i$ and $j$ indicate the initial meson-meson (meson-baryon) and final meson-meson (meson-baryon) system, respectively. 
In the on-shell ansatz, the two-particle propagators---often called loop functions---form a diagonal matrix $G$. The loop functions are
\be \label{eq:loop} G_i (s)= i \gamma_i \int \frac{d^4k}{(2\pi)^4} \ D_H(p-k) D_l(q+k) \ee
where $s=(p+q)^2$ and $D_H$ and $D_l$ are the propagators of the heavy and light particles, respectively, in the channel $i$. The
factor $\gamma_i$ accounts for the different normalization of the meson-meson and meson-baryon
interactions ($\gamma_i=1$ for the adimensional meson-meson $V$ kernel~\cite{Abreu:2011ic} while for the meson-baryon
sector $\gamma_i = 2 M_i$, with $M_i$ being the mass of the baryon~\cite{GarciaRecio:2008dp, Gamermann:2010zz, Romanets:2012hm,GarciaRecio:2012db,Garcia-Recio:2013gaa,Tolos:2013gta}). 
The loop functions are divergent and regularized by means of dimensional regularization~\cite{GarciaRecio:2008dp, Gamermann:2010zz, Abreu:2011ic, Romanets:2012hm,GarciaRecio:2012db,Garcia-Recio:2013gaa,Tolos:2013gta}.

Once the scattering amplitudes $T_{ij}$ are computed, the invariant matrix elements $\mathcal{M}_{ij}$ are given by
\ba
\mathcal{M}_{ij} (\sqrt{s})& = & \gamma_i^{1/2} \gamma_j^{1/2} \ T_{ij} (\sqrt{s}) \ . 
\ea

\begin{figure*}[htp]
\begin{center}
\includegraphics[scale=0.4]{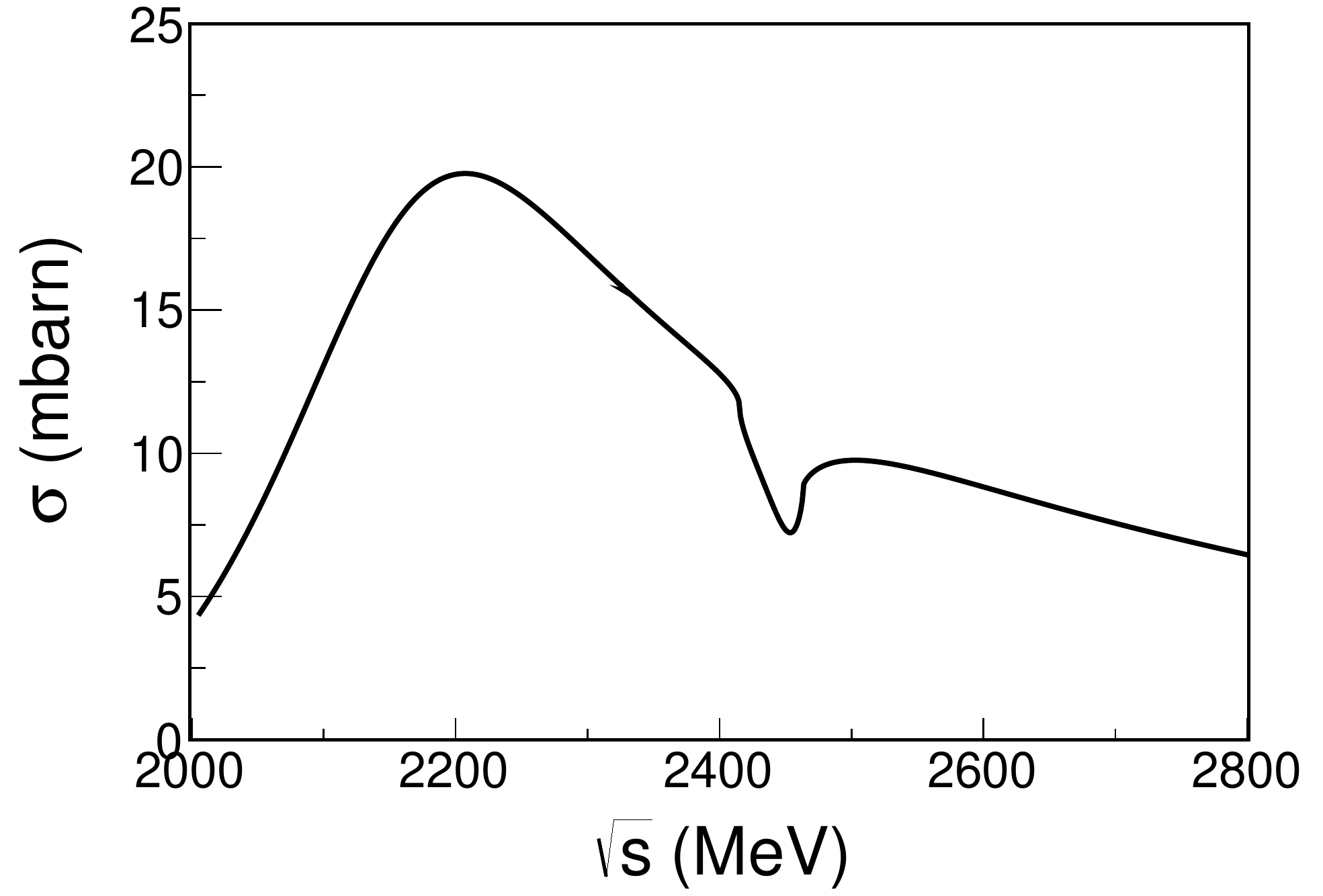}
\includegraphics[scale=0.4]{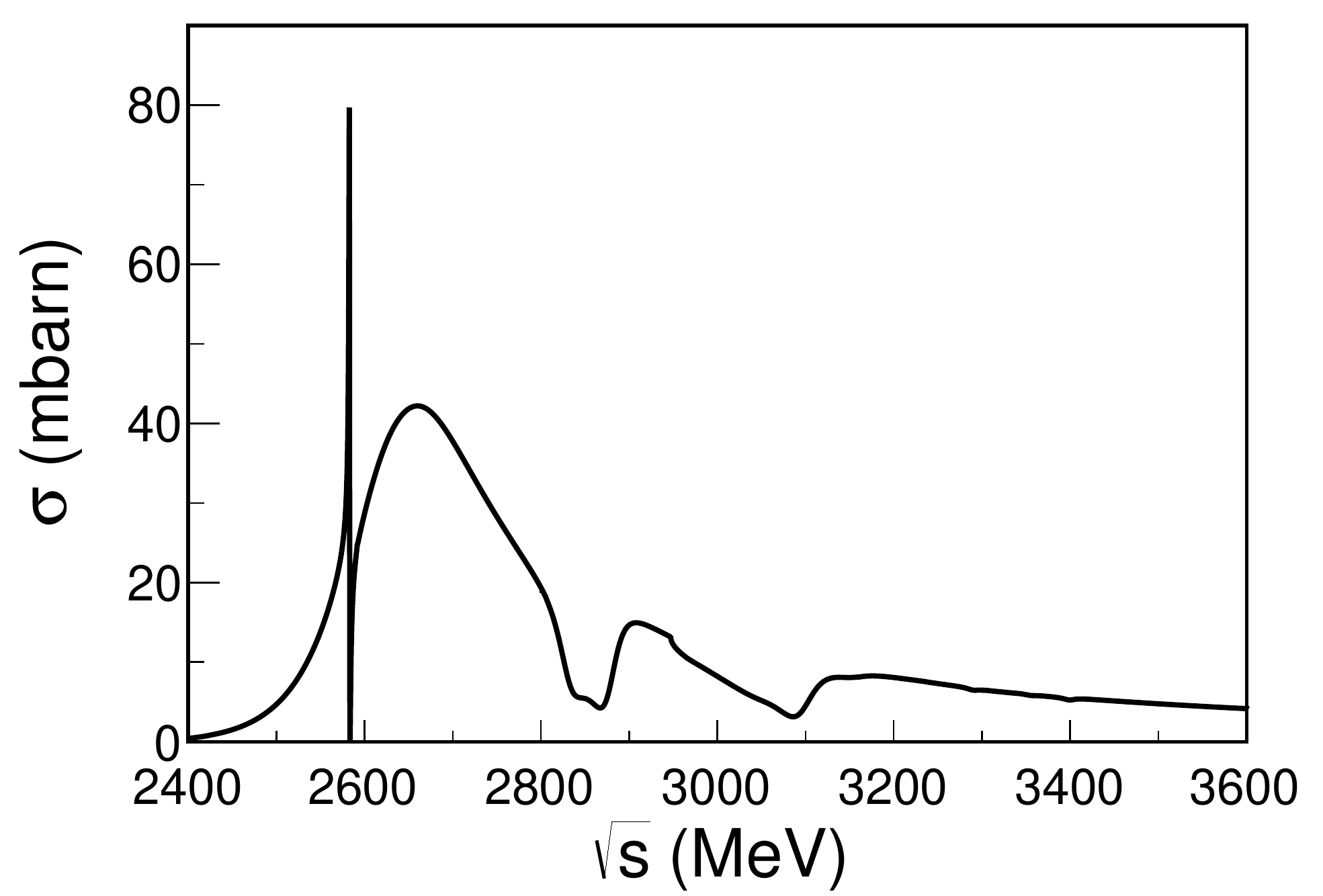}
\caption{\label{fig:charmCrossSections} Elastic cross sections for $D\pi \rightarrow D\pi$ (left panel) and $\Lambda_c \pi \rightarrow \Lambda_c \pi$ (right panel) as a function of energy in the center-of-mass frame. }
\end{center} 
\end{figure*}

\begin{figure*}[htp]
\begin{center}
\includegraphics[scale=0.4]{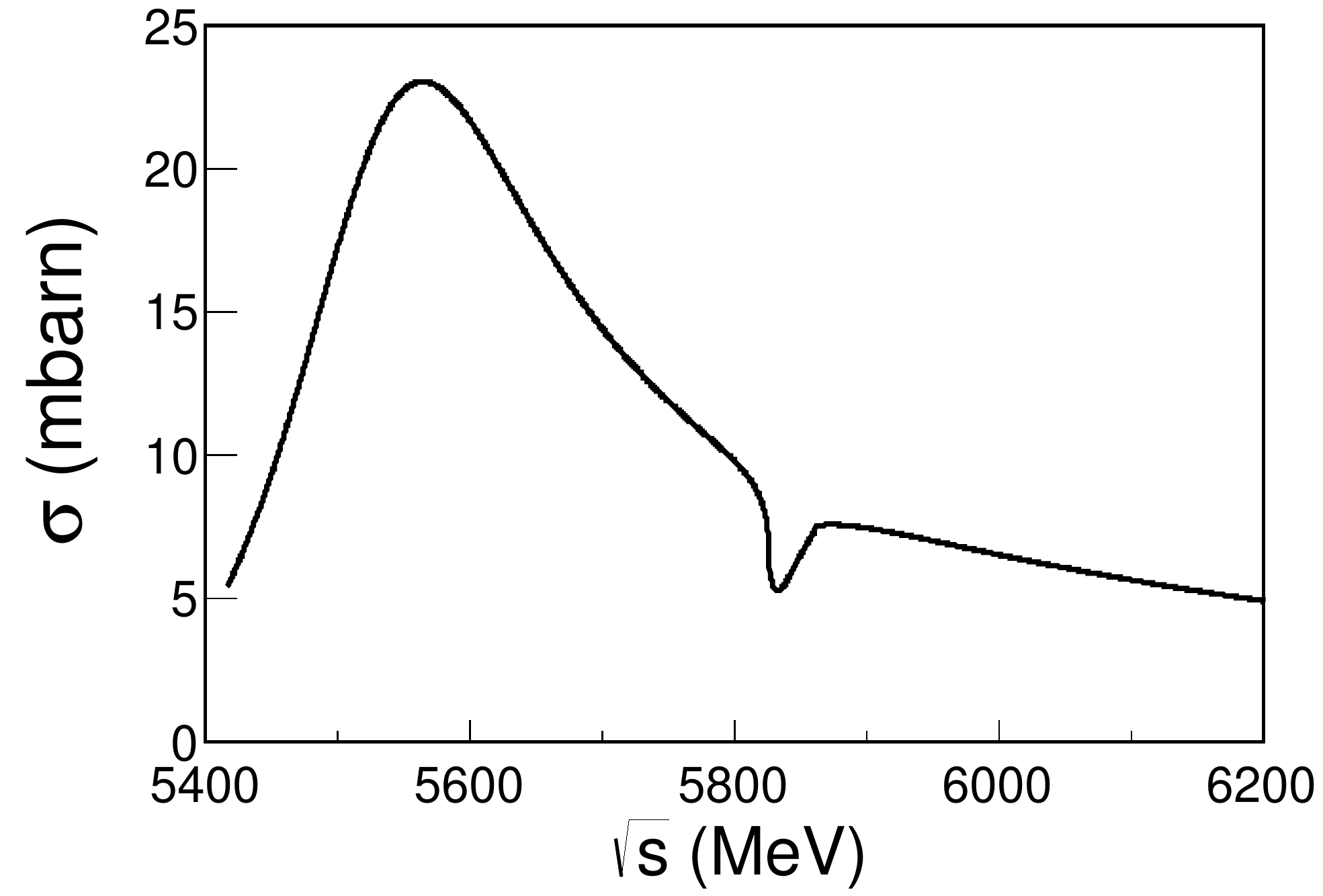}
\includegraphics[scale=0.4]{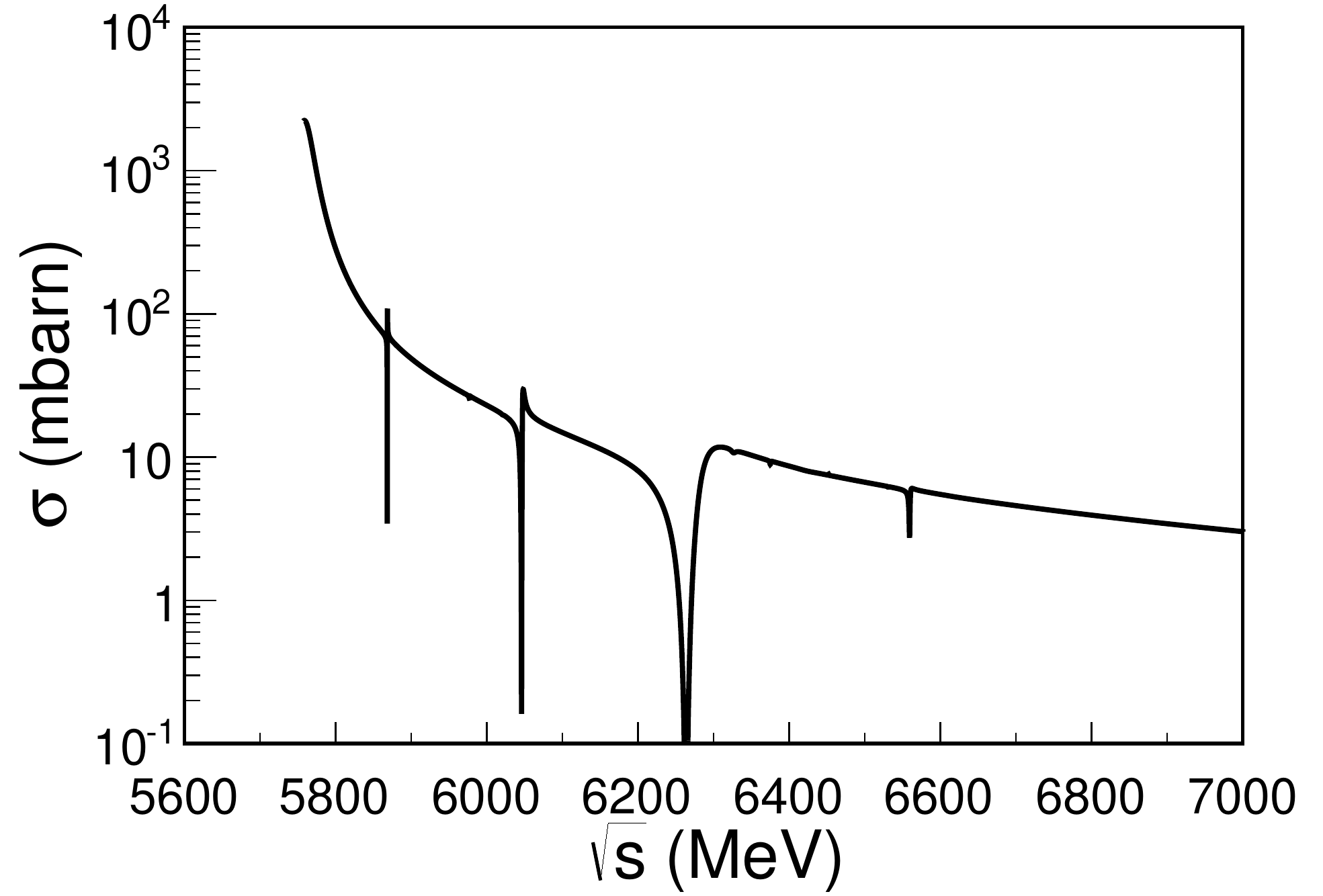}
\caption{\label{fig:bottomCrossSections} Elastic cross sections for ${\bar B} \pi \rightarrow {\bar B} \pi$ (left panel) and $\Lambda_b \pi \rightarrow \Lambda_b \pi$ (right panel)  as a function of energy in the center-of-mass frame.}
\end{center} 
\end{figure*}

\subsection{Elastic cross sections of heavy hadrons with pions}

Here we present a few results on elastic cross sections of the heavy hadrons with the most abundant light mesons
in the thermal bath, the pions. We provide these examples to show the typical features that the cross sections
present after unitarization, in particular, the resonant structures that emerge. In
Fig.~\ref{fig:charmCrossSections} we show the $D\pi$ and $\Lambda_c \pi$ elastic cross sections while
in Fig.~\ref{fig:bottomCrossSections} we display the $\bar B\pi$ and $\Lambda_b \pi$ elastic ones as a function
of energy in the center-of-mass frame. 

These elastic cross sections are obtained as 
\be
\sigma_i (s)= \frac{1}{16 \pi s}  | \mathcal{M}_{ii} (\sqrt{s}) |^2 \ .
\ee
We observe a strong energy dependence in all cross sections, given by the appearance of several resonant states that
couple strongly to these meson-baryon systems. These are described in Refs.~\cite{Romanets:2012hm,GarciaRecio:2012db,Tolos:2013kva,Torres-Rincon:2014ffa}. 
The $D\pi$ and $\bar B\pi$ cross sections are qualitatively similar, that is, a broad resonance with a characteristic
peak and a dip due to the opening of the $D\eta$ ($\bar{B}\eta)$ channel. This is due to the fact that the kernel of the interaction, $V$ in
Eq.~(\ref{eq:potm}), is the same for both systems, except for the slightly different values of $h_i$. The
main difference between both cross sections arises in the unitarization procedure due to the different masses of $D$ and $\bar B$. However, we find that the cross sections
for $\Lambda_c \pi$ and $\Lambda_b \pi$ are not comparable. The 
interaction kernels for $\Lambda_c \pi$--$\Lambda_c \pi$ and $\Lambda_b \pi$--$\Lambda_b \pi$ are quite different due to the different masses of the $\Lambda_c$ and $\Lambda_b$, as seen in Eq.~(\ref{eq:potb}), even before unitarization.

\section{Transport coefficients of $\Lambda_c$ and $\Lambda_b$ baryons \label{sec:transport}}

   In this section we use the unitarized scattering amplitudes computed from effective theories in Sec.~\ref{sec:EFT} to calculate 
the transport coefficients of the heavy baryons $\Lambda_c$ and $\Lambda_b$ in a thermal bath of light mesons.
For an ulterior application to high-energetic heavy-ion collisions at LHC and RHIC, we always consider vanishing 
net baryonic density by setting the chemical potential $\mu_B=0$.

\subsection{The transport equation for a heavy baryon}

We consider the elastic collision of a heavy baryon (denoted by $H$) with a light particle (denoted by $l$), as 
seen in Eq.~(\ref{eq:collision}). In the following we remove the particle labels $H$ and $l$ 
(unless needed for clarity), so that the particle species are specified by their momenta.
For example, when denoting any distribution function or energy by a subindex $p$ or $p-k$, it should be understood
that it refers to the heavy particle ($\Lambda_c,\Lambda_b$), which follows Fermi statistics. On the contrary, if
the distribution function or energy carry a $q$ or $q+k$ label, they refer to any of the light 
particles ($\pi,K,{\bar K},\eta$), which follows Bose statistics.

The Boltzmann equation for the distribution function of the heavy particle with momentum $p$, $f_p$ (in the absence of external forces) reads~\cite{Torres-Rincon:2012sda}
\be \label{eq:BUU} \frac{df_p}{dt} = g_l \int_{k,q} d\Gamma_{p,q \rightarrow p-k,q+k} \ [ \ f_{p-k} f_{q+k}  (1 - f_p) (1 + f_q ) 
 - f_p f_q (1 - f_{p-k}) (1 + f_{q+k}) \ ] \ , \ee
where $f_q$ is the distribution function of the light particle with momentum $q$ and $g_l$ is the spin-isospin
degeneracy factor for the light particle. Notice that in the right-hand side one should formally consider a sum over all the light 
mesons in the bath. However, we omit the summation and write our equation for a single light species for simplicity.

 The quantity $d\Gamma$ is given by
\be \label{eq:dGamma} d\Gamma_{p,q \rightarrow p-k,q+k} \equiv  \frac{d {\bf q}}{(2\pi)^3} \frac{d {\bf k}}{(2\pi)^3} 
\frac{1}{2E_p 2E_q 2E_{p-k} 2E_{q+k}} \ 2\pi  \delta (E_p+E_q-E_{p-k}-E_{q+k}) \ \overline{|T|^2} (p,q,k) \ , \ee
with the property $d\Gamma_{p,q \rightarrow p-k,q+k} = d\Gamma_{p-k,q+k \rightarrow q,p}$.

We assume that light particles are already in equilibrium as opposed to the heavy particles. Therefore,
$f_q$ (and $f_{q+k}$) will be equal to the equilibrium distribution function $n_q$ (and $n_{q+k}$),
\be n_q = g_l \frac{1}{e^{\beta (E_q-\mu_l)} - 1} \ ,  \  \ee
with $\mu_l$ a possible pseudo-chemical potential that we set to zero. The Boltzmann equation is then simplified
\be \label{eq:be} \frac{df_p}{dt} = g_l \int_{k,q}  d\Gamma_{p+k,q \rightarrow p,q+k} f_{p+k} n_{q} (1 - f_p)
 (1 + n_{q+k})  - \int_{k,q} d\Gamma_{p,q \rightarrow p-k,q+k}  f_p n_q (1 - f_{p-k}) (1 + n_{q+k}) \ , \ee
where in the first term we have used the identity 
\be d\Gamma_{p,q \rightarrow p-k,q+k} f_{p-k} n_{q+k}  (1 - f_p) (1 + n_q) = d\Gamma_{p,q+k \rightarrow p+k,q} f_{p+k} n_{q} (1 - f_p) (1 + n_{q+k})  \ , \ee
using the fact that $\overline{|T|^2}$ only depends on its arguments through the Mandelstam variables ($s,t$ and $u$).

To simplify the notation of the kinetic equation it is interesting to introduce the so-called interaction rate
$w({\bf p},{\bf k})$, which is given by~\cite{Svetitsky:1987gq,Abreu:2011ic}
\be \label{eq:intrate} d {\bf k} \ w ({\bf p},{\bf k}) \equiv \frac{ g_l}{(2\pi)^3} \int_q d\Gamma_{p,q \rightarrow p-k,q+k} \ n_q (1 + n_{q+k} )  \ . \ee
We can now rewrite the Boltzmann equation in Eq.~(\ref{eq:be}) as
\be \frac{df_p}{dt} =  \int_k d {\bf k} \ [f_{p+k} (1 - f_p) w({\bf p}+{\bf k},{\bf k}) - f_p (1 - f_{p-k}) w({\bf p},{\bf k})] \ . \ee
If we simplify $1 - f_p, 1-f_{p-k} \simeq 1$ due to the scarcity of heavy particles in the bath, we can obtain the standard 
expression~\cite{Svetitsky:1987gq}
\be \label{eq:BUUnew} \frac{df_p}{dt} =  \int_k d {\bf k} \ [f_{p+k} w({\bf p}+{\bf k},{\bf k}) - f_p w({\bf p},{\bf k})] \ . \ee

This equation for $f_p(t)$ indicates that the rate of change of 
the heavy particle distribution is composed by a gain term and a loss term. The next step is to exploit the fact that the
mass of heavy particles is larger than those of the light particles and than the temperature. Then, the heavy particle only receive 
small kicks from the light particles with a tiny momentum loss. Assuming that the typical transferred momentum 
is ${\bf k} \ll {\bf p}$, one can expand the interaction rate in Eq.~(\ref{eq:BUUnew}) up to second order in derivatives.
 The resulting equation is the Fokker-Planck equation~\cite{Svetitsky:1987gq, Abreu:2011ic}
\be \frac{df_p}{dt} =  \frac{\pa}{\pa p_i} \ \left\{  F({\bf p}) p_i f_p(t) + \frac{\pa}{\pa p_j} \left[ \Gamma_0 ({\bf p}) \Delta_{ij} + \Gamma_1 ({\bf p}) \frac{p_i p_j}{p^2} \right] f_p(t) \right\} , \ee
where $\Delta_{ij}=\delta_{ij}-p_ip_j/p^2$, $F$ is the drag force, and $\Gamma_0$ and $\Gamma_1$ are the (momentum) diffusion transport coefficients
in an isotropic thermal bath. These depend on the properties of the bath, temperature and density. The explicit expressions of these coefficients in terms of $w(\mathbf{p},\mathbf{k})$ are:
\ba \label{eq:Fcoeff}
F({\bf p},T) & =& \int_k d\mathbf{k} \ w(\mathbf{p},\mathbf{k}) \frac{k_ip^i}{p^2} \ , \\
\label{eq:G0coeff} \Gamma_0({\bf p},T) & =& \frac{1}{4} \int_k d\mathbf{k} \ w(\mathbf{p},\mathbf{k}) \left[ \mathbf{k}^2
- \frac{(k_ip^i)^2}{p^2} \right] \ ,  \\
\label{eq:G1coeff} \Gamma_1( {\bf p},T) & =& \frac{1}{2} \int_k d\mathbf{k} \ w(\mathbf{p},\mathbf{k}) \frac{(k_ip^i)^2}{p^2} \ . 
\ea
These integrals can be computed using Monte Carlo techniques. Details on the computation are given in Ref.~\cite{Abreu:2011ic}.

In the so-called static limit, ${\bf p} \rightarrow 0$, the coefficients in Eqs.~(\ref{eq:Fcoeff},\ref{eq:G0coeff},\ref{eq:G1coeff}) 
follow simple relations. In particular the diffusion coefficients become degenerate $\Gamma \equiv  \Gamma_0({\bf p} \rightarrow 0)
= \Gamma_1 ({\bf p}\rightarrow 0)$ and are related to the drag force through the Einstein relation
\be \label{eq:einstein} F \equiv F( {\bf p} \rightarrow 0) = \frac{\Gamma}{m_H T} \ , \ee
where $m_H$ is the mass of the heavy baryon. This relation has been checked in Ref.~\cite{Abreu:2012et} up to a good degree of accuracy, with
small violations at moderate temperatures due to the presence of the UV-momentum cutoff in our integrals. Notice that this cutoff in the 
integrations~(\ref{eq:Fcoeff},\ref{eq:G0coeff},\ref{eq:G1coeff}) is mandatory, as one is not legitimate to use the results from a low-energy effective theory beyond certain momentum transfer.
In principle, one does not expect large truncation uncertainties, because thermal distributions suppress the integrand at higher momenta. However, at moderate temperatures one might
find non-negligible systematic errors if the cutoff is kept.

\subsection{Drag and diffusion coefficients from the Fokker-Planck equation}

We start this section by analyzing the transport coefficients of the $\Lambda_c$ baryon. We consider a light-meson bath composed by $\pi,K,\bar{K}$ and $\eta$ mesons, so we fix the baryochemical potential to zero neglecting
the effects of nucleons and $\Delta$ baryons on the transport of the $\Lambda_c$. This approach is valid for high energetic heavy-ion collisions, such as those taking place at LHC or RHIC colliders at their top energies.

We follow a similar procedure as in our last work for $D$ mesons~\cite{Tolos:2013kva}. However, in contrast to the calculation in Ref.~\cite{Tolos:2013kva}, we remove the momentum UV-momentum cutoff in the integrals
of Eqs.~(\ref{eq:Fcoeff},\ref{eq:G0coeff},\ref{eq:G1coeff}) by taking constant cross sections for energies above the validity of the effective theory. This procedure eliminates the systematic
uncertainty of previous calculations due to the truncation of the momentum integrals, but introduces an uncertainty---especially at high temperatures---due to the use of a constant interaction at high energies. However,
we prefer to pursue this path (inspired by Regge analysis at moderate energies, where one finds almost flat cross sections) as we eventually need to go to high values of transverse momentum in our calculations of heavy-ion observables at RHIC and LHC energies~\cite{Das:2016llg}.

\begin{figure*}[htp]
\begin{center}
\includegraphics[scale=0.4]{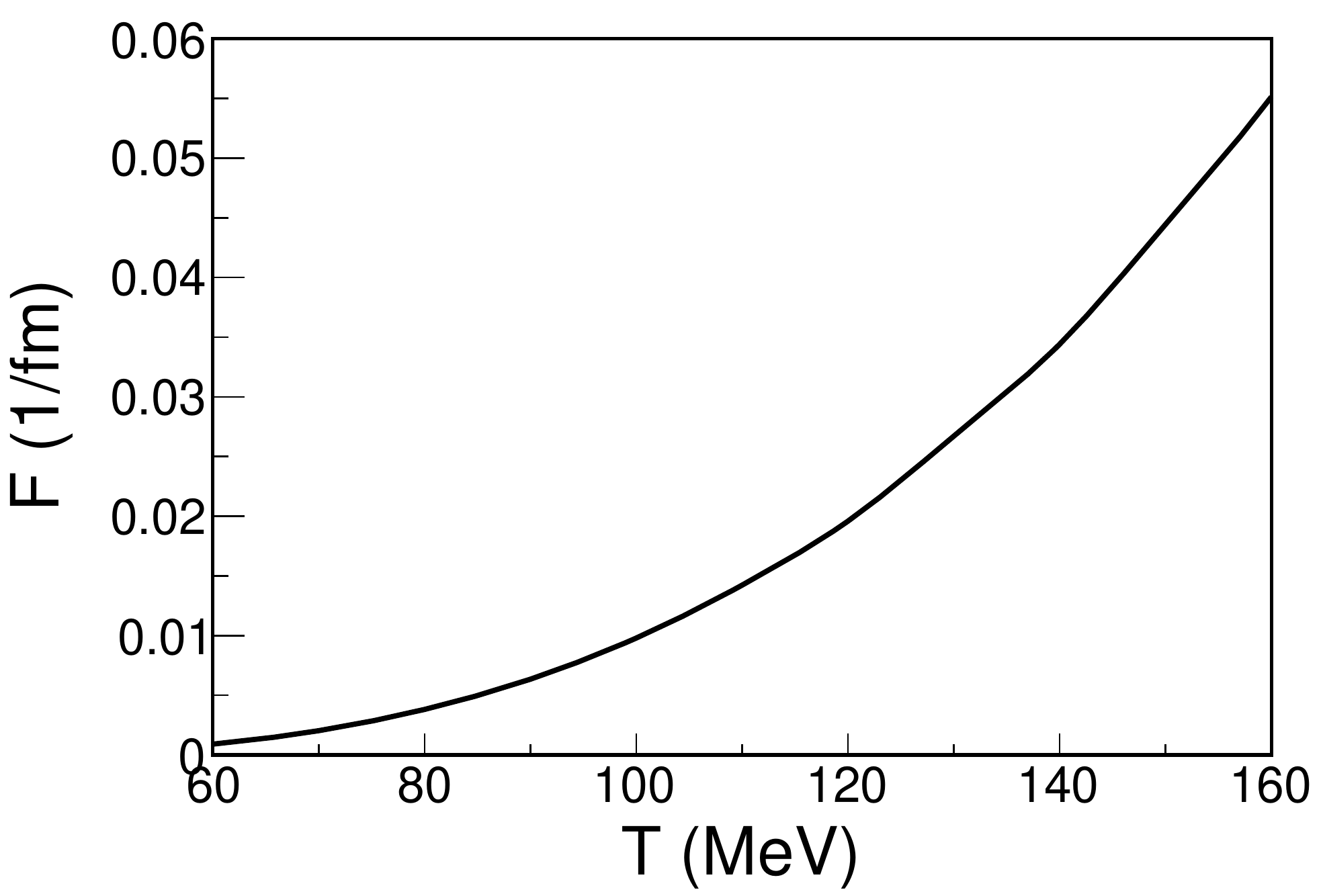}
\includegraphics[scale=0.4]{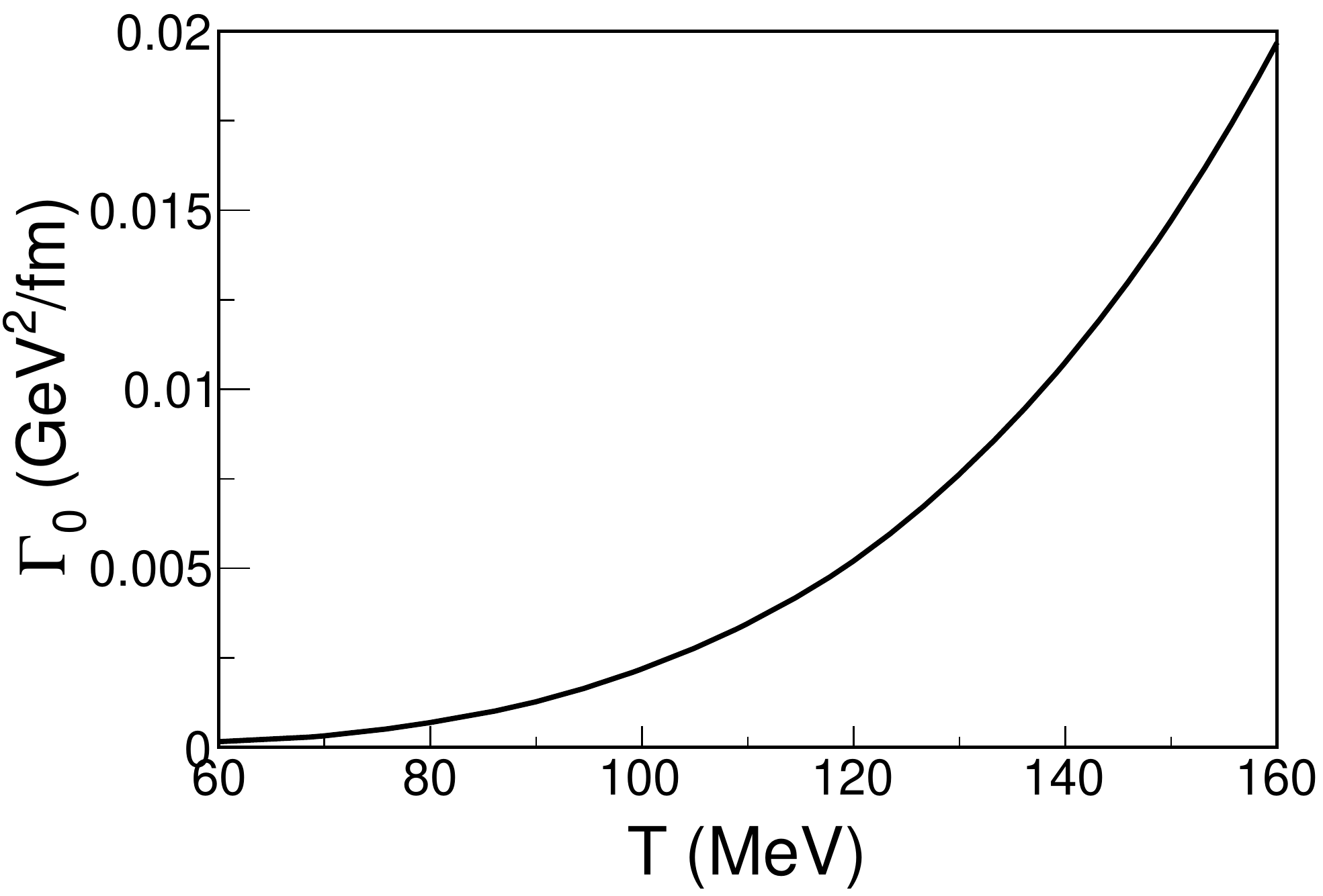}
\caption{\label{fig:LambdacDrag1} Drag $F$ (left panel) and diffusion coefficient $\Gamma_0$ (right panel) in the static limit ($p=100$ MeV)
as a function of the temperature for $\Lambda_c$ baryons in a gas of thermalized $\pi,K,\bar{K},\eta$ mesons. }
\end{center} 
\end{figure*}

\begin{figure*}[htp]
\begin{center}
\includegraphics[scale=0.4]{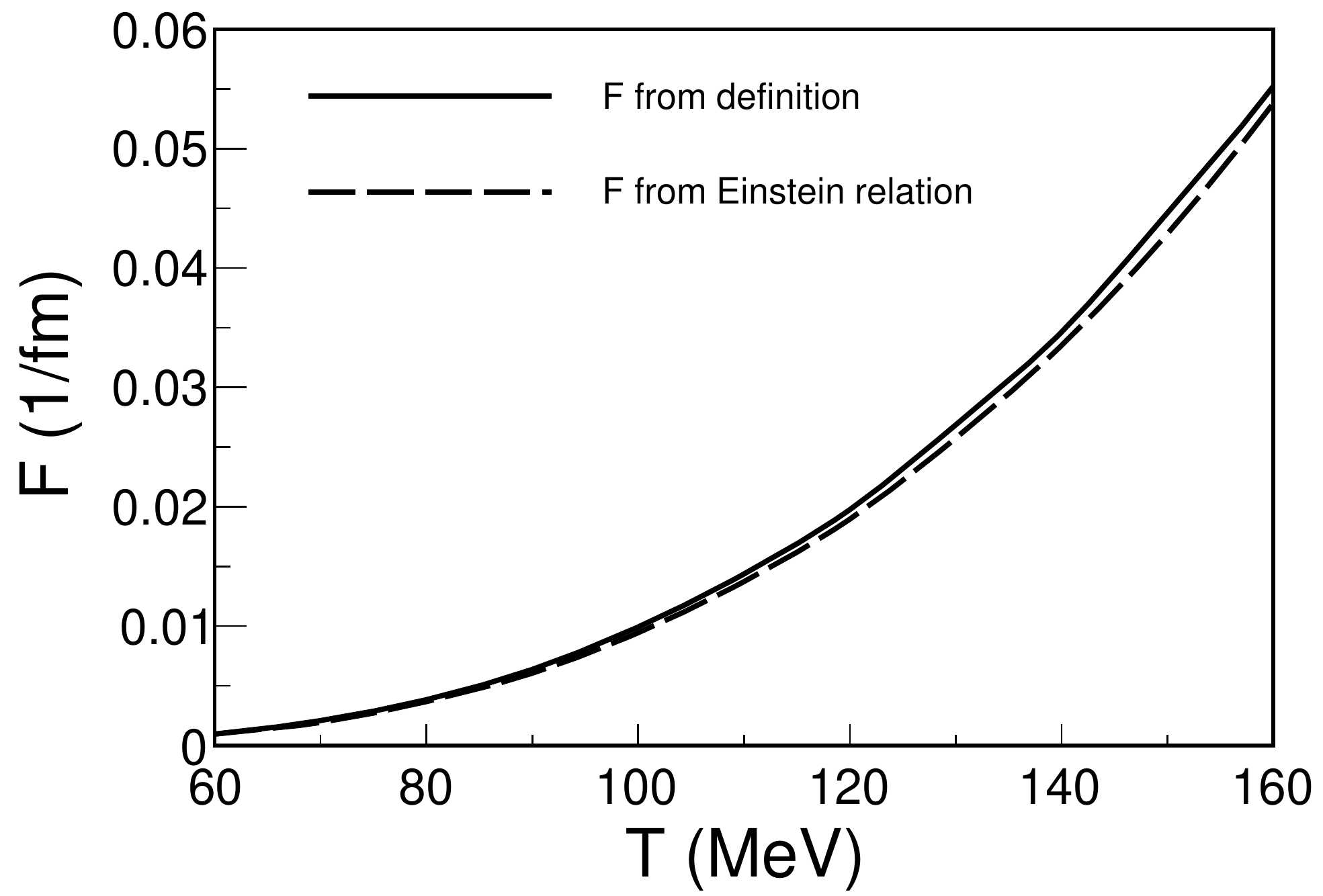}
\caption{\label{fig:LambdacDragEins} Analysis of the applicability of the Einstein relation of Eq.~(\ref{eq:einstein})
for the $\Lambda_c$ baryon. Solid line: Drag coefficient 
as computed from its definition of Eq.~(\ref{eq:Fcoeff}). Dashed line: Drag coefficient calculated
using Eq.~(\ref{eq:einstein}), with $\Gamma$ being the diffusion coefficient $\Gamma_0$ of Eq.~(\ref{eq:G0coeff}) in the static limit ($p=100$ MeV).}
\end{center} 
\end{figure*}

In Fig.~\ref{fig:LambdacDrag1}  we show the drag and diffusion coefficients for $\Lambda_c$ as a function of the temperature when the heavy-baryon momentum is fixed
to $p=100$ MeV (static limit). We plot the two coefficients as we sum up the contribution of all light mesons. Note that in this limit the two diffusion coefficients, $\Gamma_0$ and $\Gamma_1$, are degenerate, so
we only show $\Gamma_0$ in the right panel. In the static limit, the Einstein relation of Eq.~(\ref{eq:einstein}) should be satisfied. Thus, we have analyzed its applicability by evaluating the drag coefficient from Eq.~(\ref{eq:einstein})
using the diffusion coefficient calculated by Eq.~(\ref{eq:G0coeff}). The excellent agreement of the two calculations
is presented in Fig.~\ref{fig:LambdacDragEins} for a wide range of temperatures.
In our past works \cite{Tolos:2013kva,Torres-Rincon:2014ffa} we reported a small violation of the Einstein relation due to the presence of the  UV-momentum cutoff~\cite{Abreu:2012et},
being more severe at high temperatures. The removal of this cutoff dependence in the present work has improved the agreement between both approaches.

If we compare the drag coefficient for the $\Lambda_c$ in Fig.~\ref{fig:LambdacDragEins} with the one for the $D$ meson
in Ref.~\cite{Tolos:2013kva}, we find that they are very similar (this also happens for the
diffusion coefficient). This unexpected similarity can be understood in terms of the kinetic theory expressions of
the drag and diffusion coefficients in the nonrelativistic limit~\cite{lifschitz1983physical}
\ba \label{eq:estimateF}  F & \sim  & P \sigma \sqrt{ \frac{m_l}{T}} \frac{1}{m_H} \ , \\
 \label{eq:estimateG} \Gamma &\sim& P \sigma \sqrt{m_l T} \ . \ea
The pressure results from the light particle's bath, which is taken to be the same for $D$ meson and $\Lambda_c$ baryon,
with the pionic pressure being the dominant contribution. The masses of the two hadrons are comparable ($D$ meson mass is
about 80 \% the mass of the $\Lambda_c$). Moreover, the cross sections for the $D\pi$ and $\Lambda_c\pi$ scattering of Fig.~\ref{fig:charmCrossSections} 
are alike (of course, in different kinematic range), so that the thermally averaged cross sections should be quite
similar in size. Thus, one expects similar values of the drag and 
diffusion coefficients for $D$ and $\Lambda_c$ hadrons (at least in the static limit, where Eqs.~(\ref{eq:estimateF}-\ref{eq:estimateG}) are valid).

\begin{figure*}[htp]
\begin{center}
\includegraphics[scale=0.4]{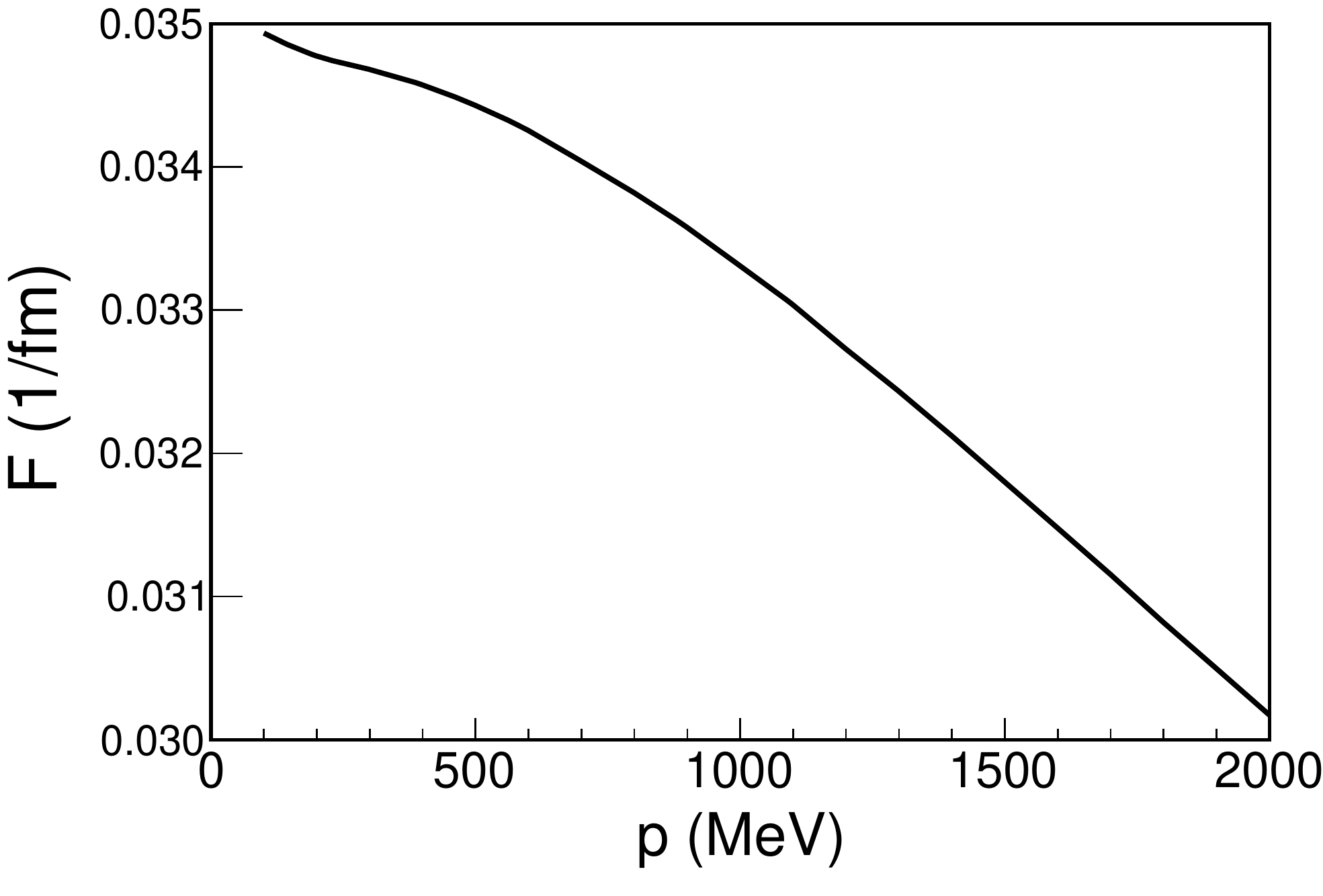}
\includegraphics[scale=0.4]{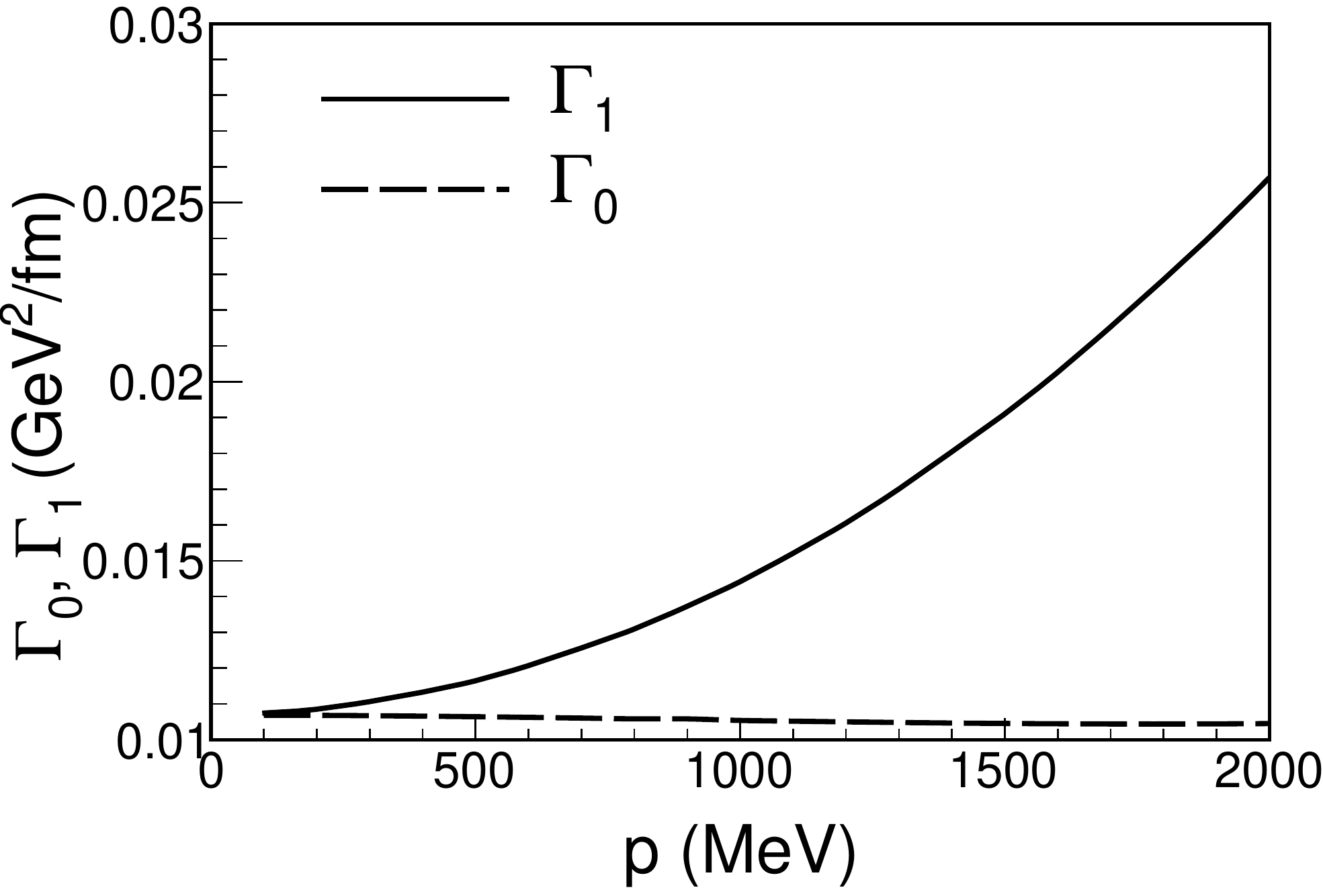}
\caption{\label{fig:LambdacDrag2} Drag (left panel) and diffusion (right panel) coefficients at $T=140$ MeV as a function of the $\Lambda_c$
momentum in a gas of thermalized $\pi,K,\bar{K},\eta$ mesons. }
\end{center} 
\end{figure*}

The momentum dependence of the $\Lambda_c$ transport coefficients is shown in Fig.~\ref{fig:LambdacDrag2}, where we 
plot the drag force and the two diffusion coefficients for $T=140$ MeV as a function of $\Lambda_c$ momentum.
This temperature is chosen to be close to the transition temperature but still in the hadronic phase.

The drag and diffusion coefficients for $\Lambda_c$ as a function of temperature were also analyzed in Ref.~\cite{Ghosh:2014oia}. 
In this work the transport coefficients were obtained by means of two approaches, using the scattering lengths of the 
different $\Lambda_c$-light meson system or the energy-dependent Born terms (or interaction kernels), without unitarization. It was found that 
the coefficients are smaller in size than the ones reported in the present paper. Apart from differences in the kernel of 
the interaction, we conclude that the role of unitarization is important for the correct determination of the transport 
coefficients due to the presence of resonant states, that enhance the cross section, and therefore, the transport coefficients themselves. 

\begin{figure*}[htp]
\begin{center}
\includegraphics[scale=0.4]{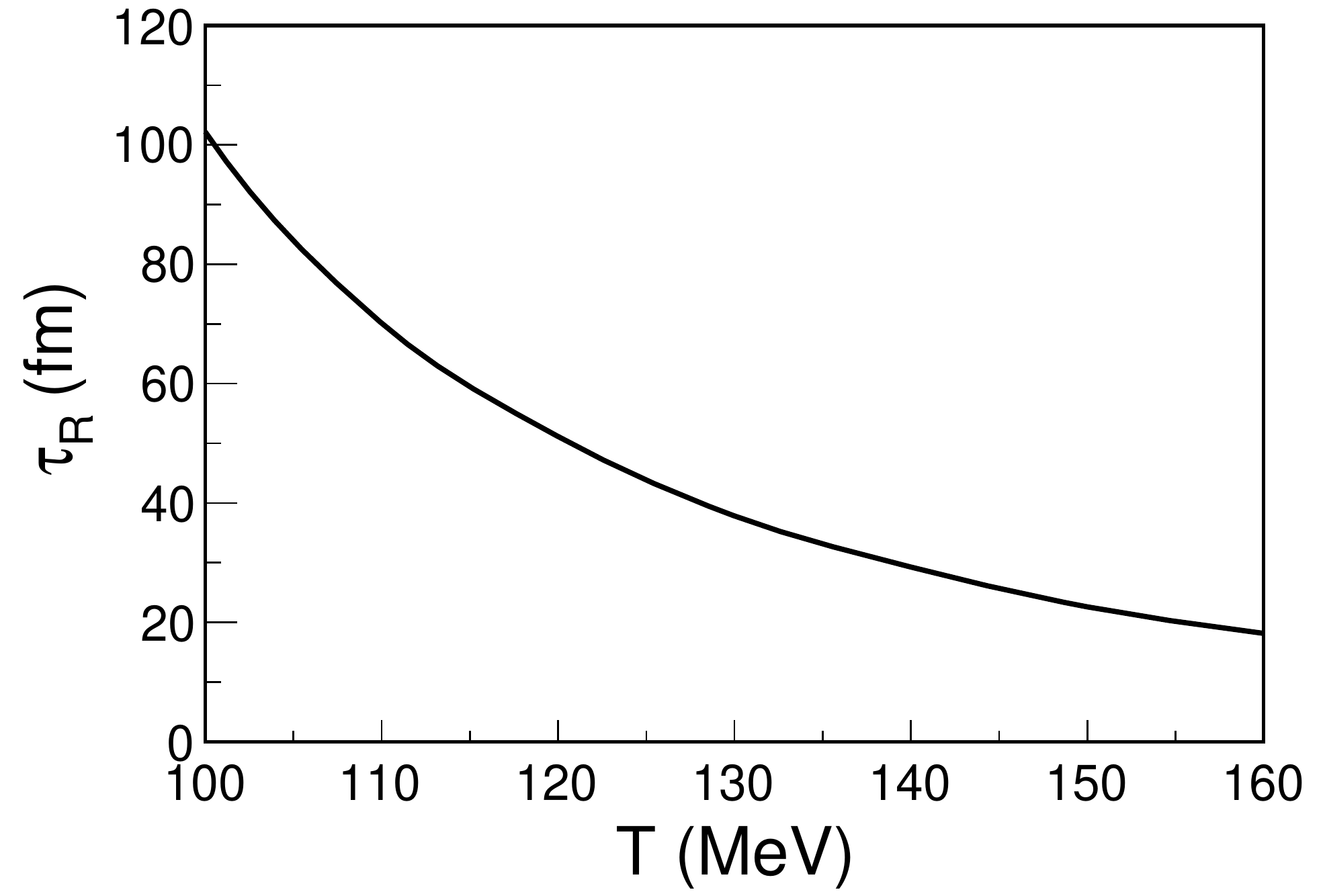}
\includegraphics[scale=0.4]{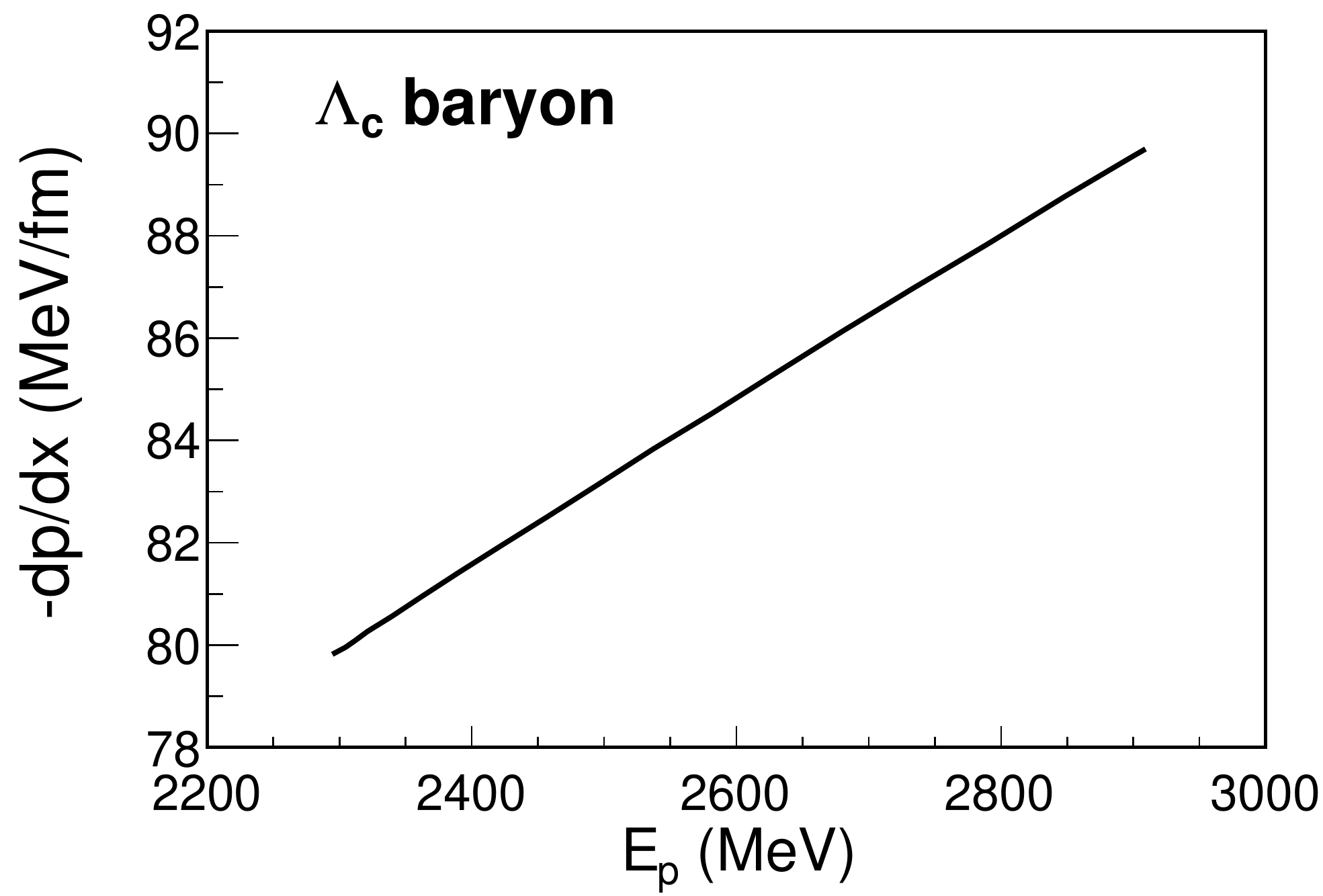}
\caption{\label{fig:Lambdactau}  
Left panel: Relaxation time of $\Lambda_c$ baryons in the static limit $p\rightarrow 0$ as a function of the temperature. Right panel: Average momentum loss of $\Lambda_c$ baryons in a thermal medium at $T=140$ MeV 
as a function of the baryon energy.
}
\end{center} 
\end{figure*}

 To conclude the analysis of the $\Lambda_c$ transport coefficients we provide two related quantities. In Fig.~\ref{fig:Lambdactau} we show 
the relaxation time for the average momentum $\tau_R=1/F$ in the static limit as a function of the temperature, and the averaged momentum loss per unit length
$-dp/dx = E_p F$ as a function of the heavy baryon energy at $T=140$ MeV. Notice that the relaxation time is larger than the typical fireball duration, which is a signature
of the difficult equilibration of heavy particles in heavy-ion collisions. When heavy particles are implemented in heavy-ion simulations, the large relaxation time is reflected
in a small number of collisions (the mean-free path also becomes of the order of the fireball duration). As an example, in Ref.~\cite{Song:2015sfa} it is quoted that each $D$ meson
can have $1-2$ collisions with light mesons in central $Au+Au$ collision at RHIC energies. In view of our similar cross sections, the same can be expected for heavy baryons.
The momentum loss is around 85 MeV per Fermi, again signalling a not very effective process of thermalization by momentum loss.
 
\begin{figure*}[htp]
\begin{center}
\includegraphics[scale=0.4]{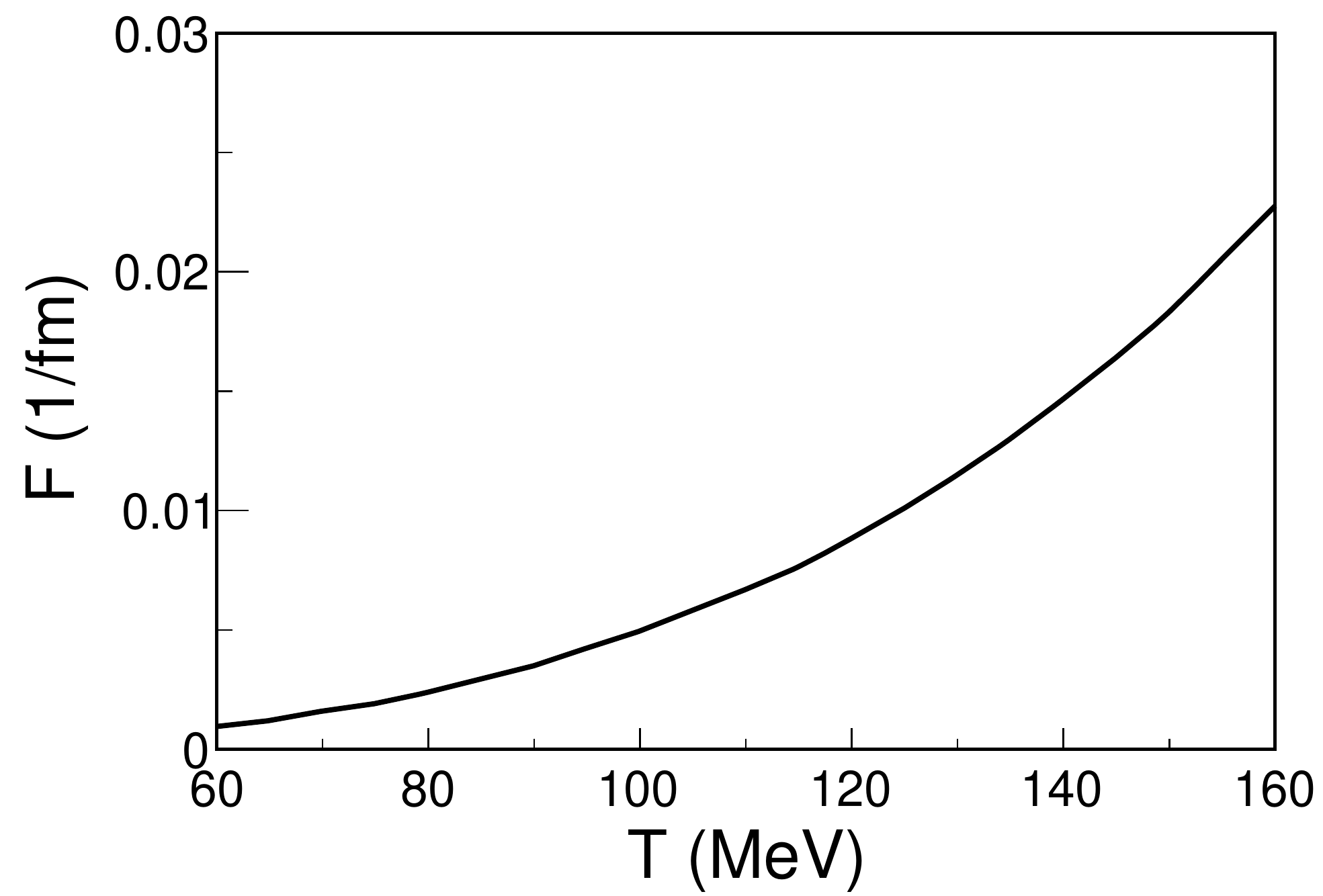}
\includegraphics[scale=0.4]{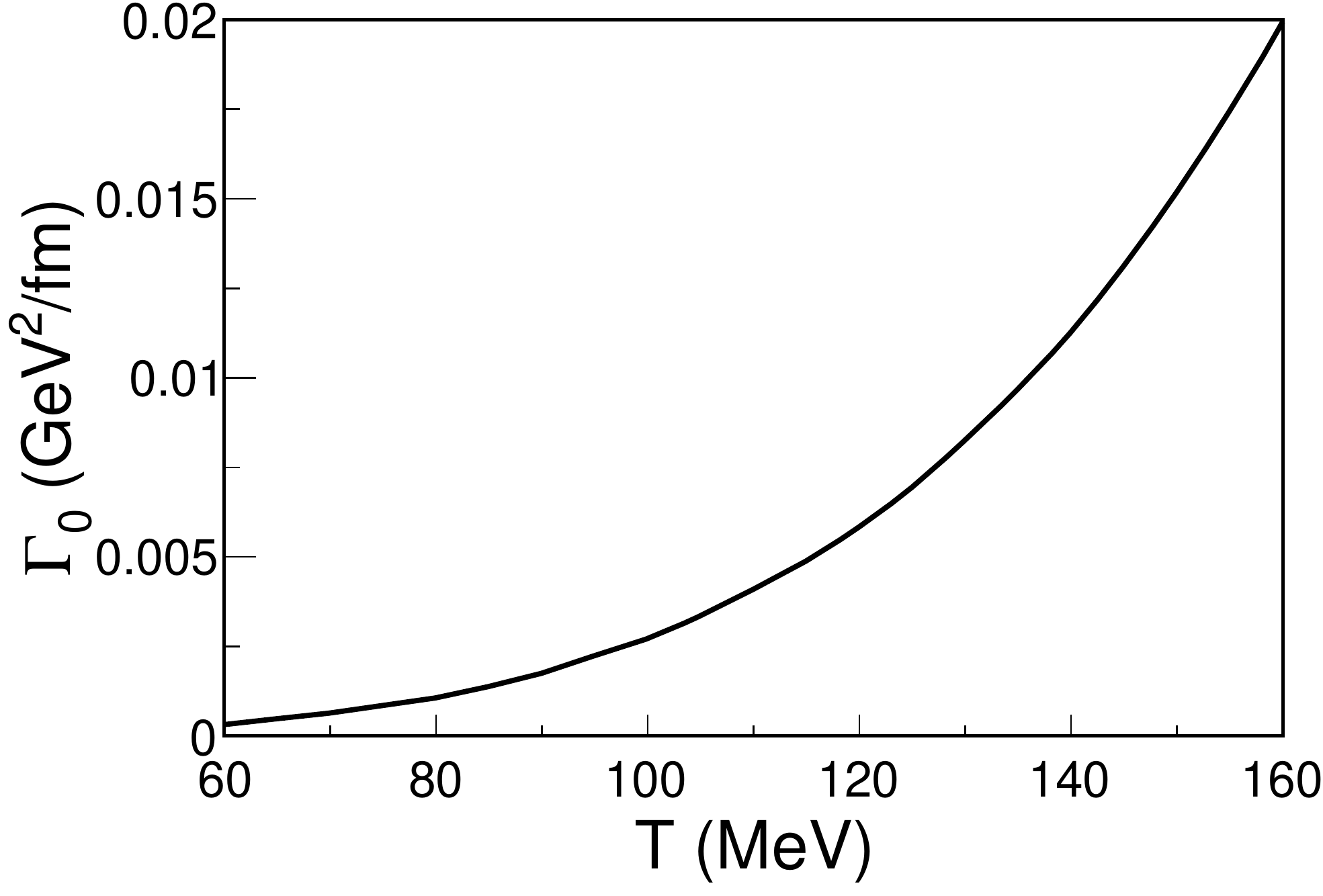}
\caption{\label{fig:F_Lambdab} 
 As in Fig.~\ref{fig:LambdacDrag1} for $\Lambda_b$. 
 }
\end{center} 
\end{figure*}

\begin{figure*}[htp]
\begin{center}
\includegraphics[scale=0.4]{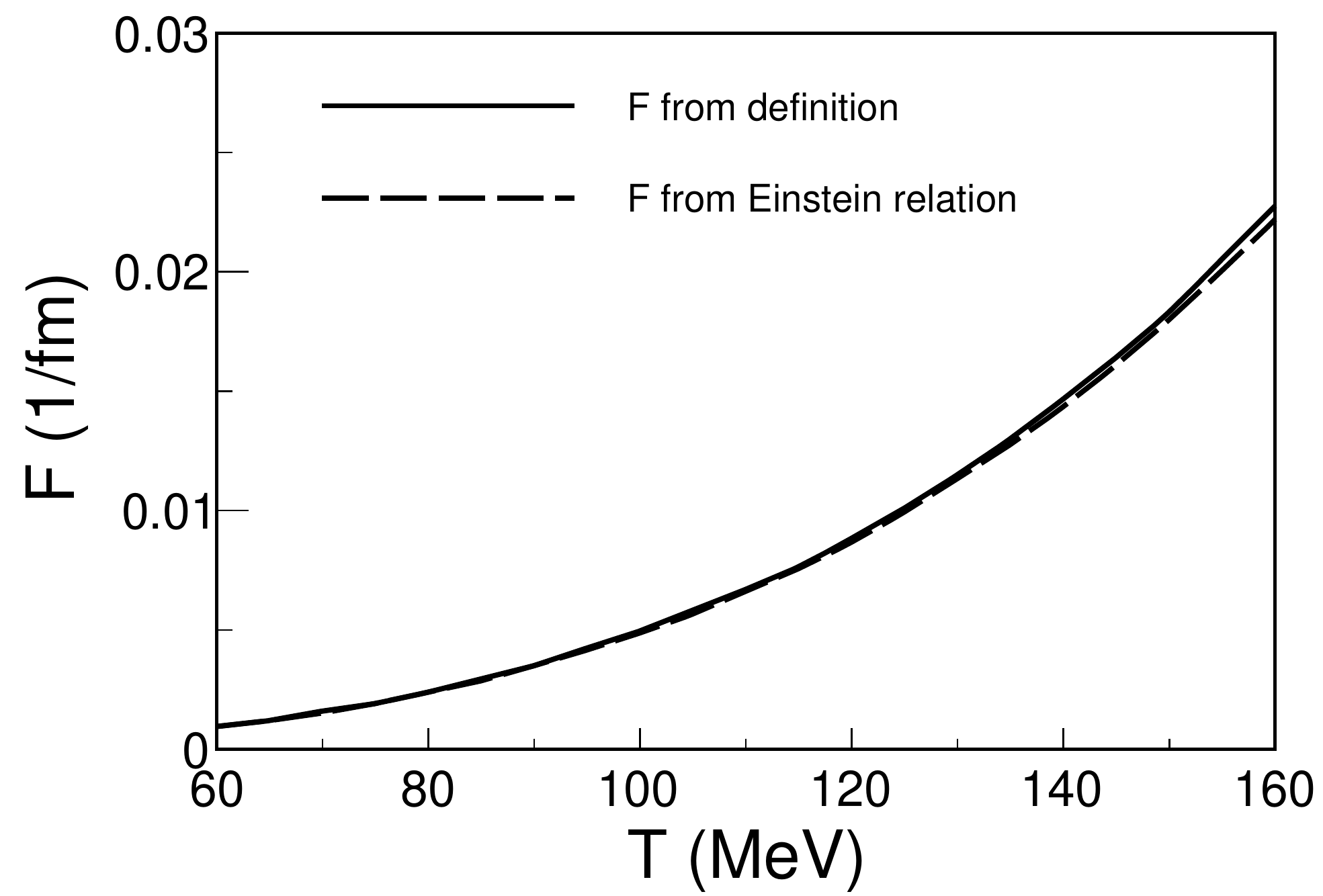}
\caption{\label{fig:LambdabDragEins} 
As in Fig.~\ref{fig:LambdacDragEins} for $\Lambda_b$.
}
\end{center} 
\end{figure*}

\begin{figure*}[htp]
\begin{center}
\includegraphics[scale=0.4]{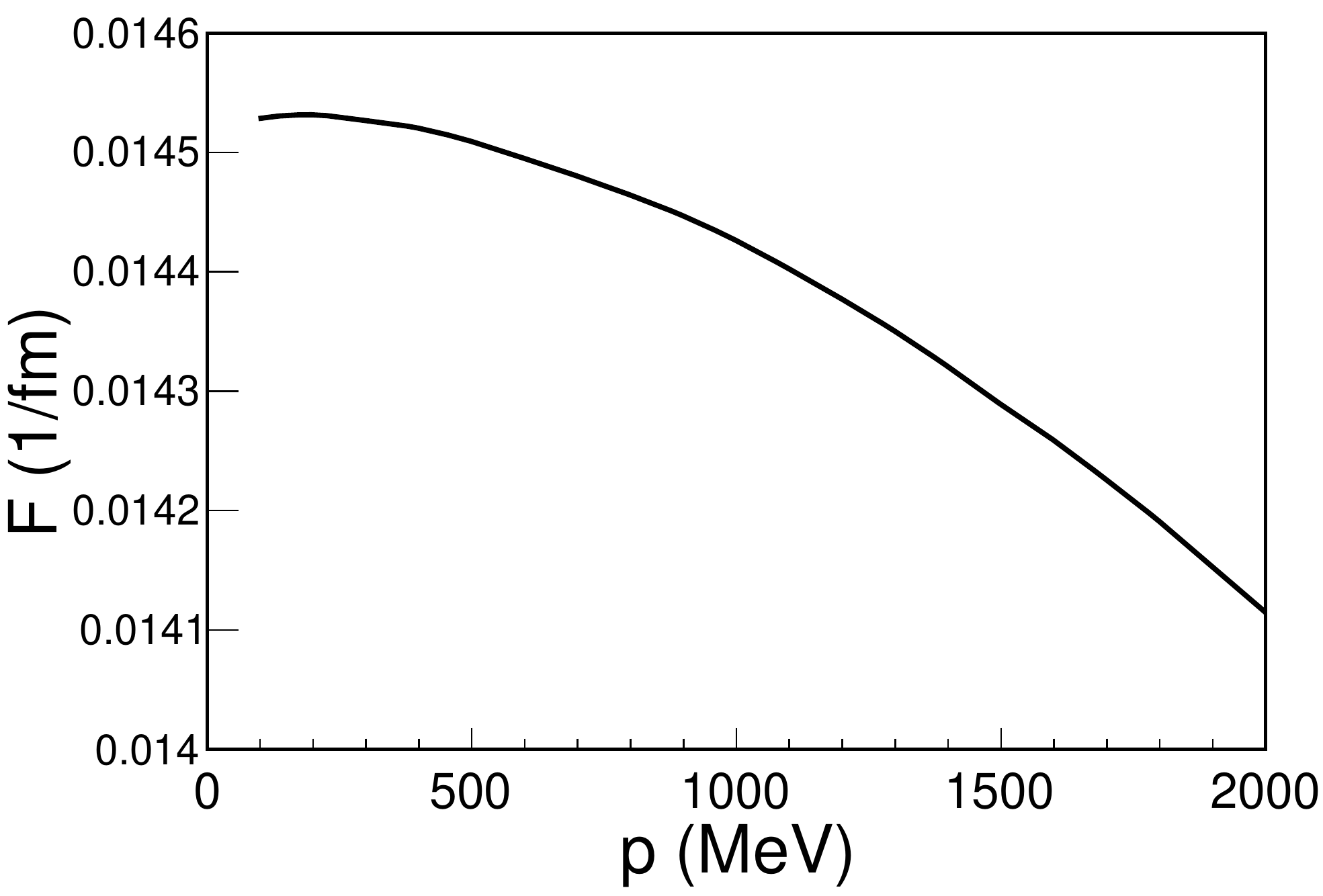}
\includegraphics[scale=0.4]{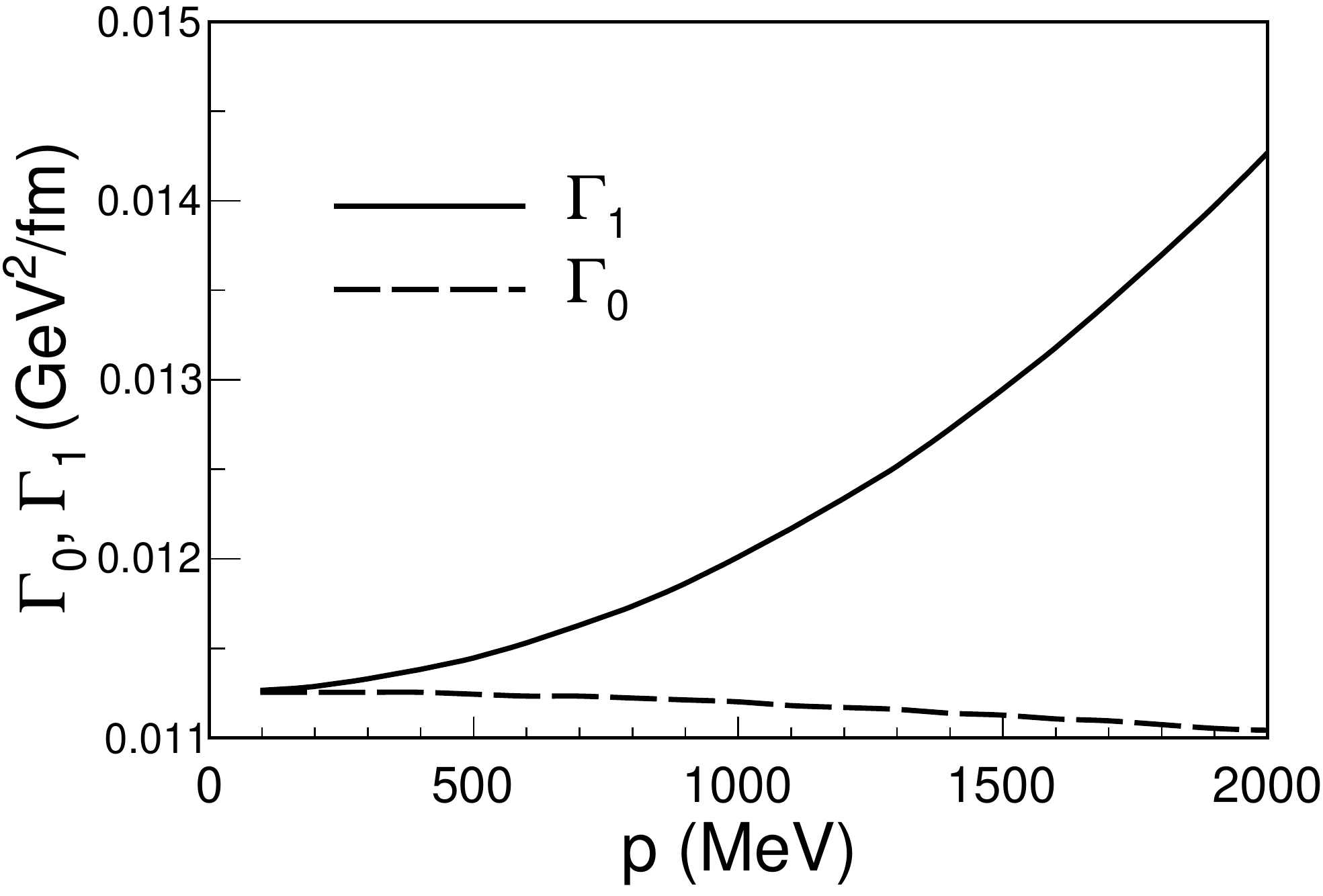}
\caption{\label{fig:Lambdab2}  
As in Fig.~\ref{fig:LambdacDrag2} for $\Lambda_b$.
}
\end{center} 
\end{figure*}

With regards to the $\Lambda_b$ baryon, we proceed analogously  to the $\Lambda_c$ baryon and study the
transport coefficients in the static limit together with the applicability of the Einstein relation. The
results for $p=100$ MeV are presented in Fig.~\ref{fig:F_Lambdab} for the same composition of the thermal bath.
The Einstein relation in Fig.~\ref{fig:LambdabDragEins} is very well satisfied for $\Lambda_b$ as we remove 
the UV-momentum cutoff of the integrals. Moreover, we show the transport coefficients as a function 
of momentum in Fig.~\ref{fig:Lambdab2}, where similar trends to those for the $\Lambda_c$ baryon are found in all the transport coefficients.

\begin{figure*}[htp]
\begin{center}
\includegraphics[scale=0.4]{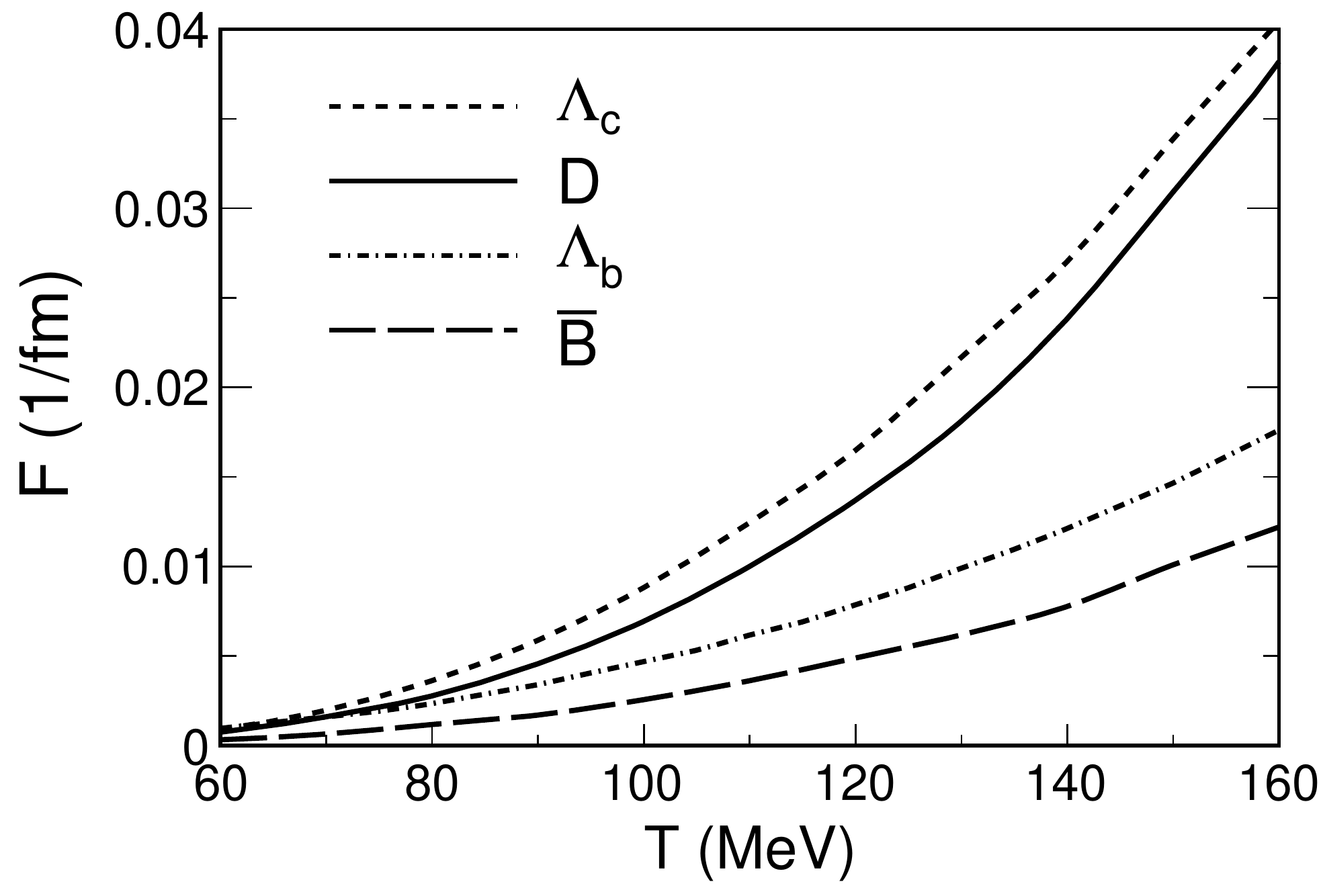}
\includegraphics[scale=0.4]{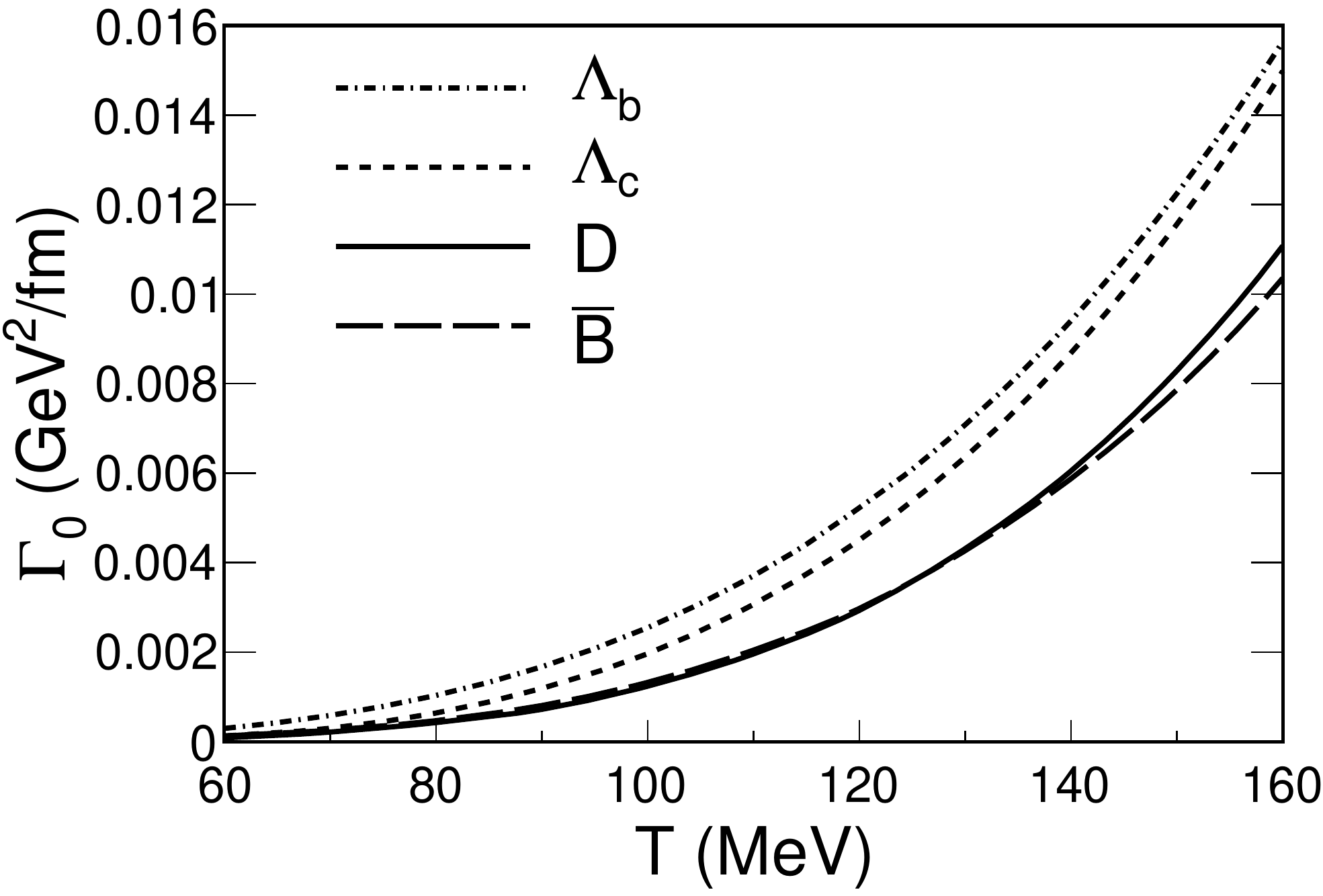}
\caption{\label{fig:allstates}  Drag (left panel) and diffusion (right panel) coefficients at $p=100$ MeV as a function
of the temperature for $D,\bar{B},\Lambda_c,\Lambda_b$ hadrons in a gas of thermalized $\pi$ mesons. }
\end{center} 
\end{figure*}

 To conclude this section we make an analysis of the drag and diffusion coefficients in the static limit, for heavy mesons and baryons, and compare them 
with the nonrelativistic estimates of Eqs.~(\ref{eq:estimateF},\ref{eq:estimateG}).  To simplify the discussion we consider a thermal bath composed only by pions and make use of the cross 
sections shown in Figs.~\ref{fig:charmCrossSections} and \ref{fig:bottomCrossSections}. We will argue that there is full consistency between the nonrelativistic estimates, the cross sections,
and the transport coefficients obtained here. We present the results for $F$ and $\Gamma_0$ in Fig.~\ref{fig:allstates}.

   We first focus on the $\Gamma_0$ coefficient (right panel). According to Eq.~(\ref{eq:estimateG}) it is only sensitive to the thermally averaged cross section of the elastic collision.

\begin{itemize}
 \item We take the $D$-meson result as reference (solid line). Assuming that Eq.~(\ref{eq:estimateG}) is valid, we expect that the difference between the $\Gamma_0$ coefficients  for the $D$ and $\bar{B}$ mesons
will only depend on the differences between the cross sections. By analyzing the $D-\pi$ and $\bar{B}-\pi$ cross sections in Figs.~\ref{fig:charmCrossSections} and \ref{fig:bottomCrossSections}, we
observe that on average they are of the same order. Therefore one can 
expect similar  $\Gamma_0$ coefficients, as seen in the right panel of Fig.~\ref{fig:allstates}.

\item Turning to the baryon case, the $\Lambda_c-\pi$ cross section has a similar qualitative trend but it is slightly larger than the $D-\pi$ one. The ratio of the average
cross sections is expected to be larger than 1 (it is a factor of 2 right at the resonance peak). Reading the ratio of the diffusion 
coefficients from the figure, which is around $\sim 1.4$, and assuming that the estimate of Eq.~(\ref{eq:estimateG}) is
valid, we take this numerical value as an indicator of $\sigma_{\Lambda_c-\pi}/\sigma_{D-\pi}$ ratio.

\item Finally, to extract any conclusions from the complicated structure of the $\Lambda_b-\pi$ cross section, one would need 
to perform a numerical computation of the average cross section. 
However, from the right panel of Fig.~\ref{fig:allstates} we observe that the result for $\Gamma_0$ is close to the $\Lambda_c$ one. 
Therefore, assuming again the validity of Eq.~(\ref{eq:estimateG}), we conclude that the average cross section for
$\Lambda_b-\pi$ should be of the same order of the one for $\Lambda_c-\pi$.
\end{itemize}

Now, we will check the consistency of these claims by looking at the results for the drag coefficient, $F$, in the left panel 
of Fig.~\ref{fig:allstates}.

\begin{itemize}
 \item Let us take again the $D$-meson result as a reference. The nonrelativistic estimate of Eq.~(\ref{eq:estimateF}) tells us that $F$ depends not only on the averaged cross section but also on the heavy mass. 
For the $\bar{B}$ meson colliding with pions, we have seen that the elastic cross sections are similar. Therefore, the drag coefficient should scale with the inverse of heavy masses. Thus, we expect the $\bar{B}$
result to be {\it reduced} by a factor $m_B/m_D \sim 2.8$, which is fully consistent with our results in left panel of Fig.~\ref{fig:allstates}.

\item For the $\Lambda_c$ baryon one expects that the result is a combination of an increase of a factor $1.4$ due to the cross section dependence, but also a reduction of $m_{\Lambda_c}/m_D \sim 1.2$. The total effect is an {\it increase} of
a factor 1.1 with respect to $D$ meson, which is easily seen in the same panel.

\item Finally, from the previous analysis of the $\Gamma_0$ coefficient, we have seen that the average cross section of $\Lambda_b-\pi$  is of the same order of the  $\Lambda_c-\pi$ case. Therefore 
we expect the same increase of $1.4$, but a reduction of $m_{\Lambda_b}/m_D \sim 3.0$ due to heavy masses. The total effect is a {\it reduction} of a factor $2.2$. The result in the left panel of Fig.~\ref{fig:allstates} is in full accordance with this expectation.
\end{itemize}

\begin{figure*}[htp]
\begin{center}
\includegraphics[scale=0.4]{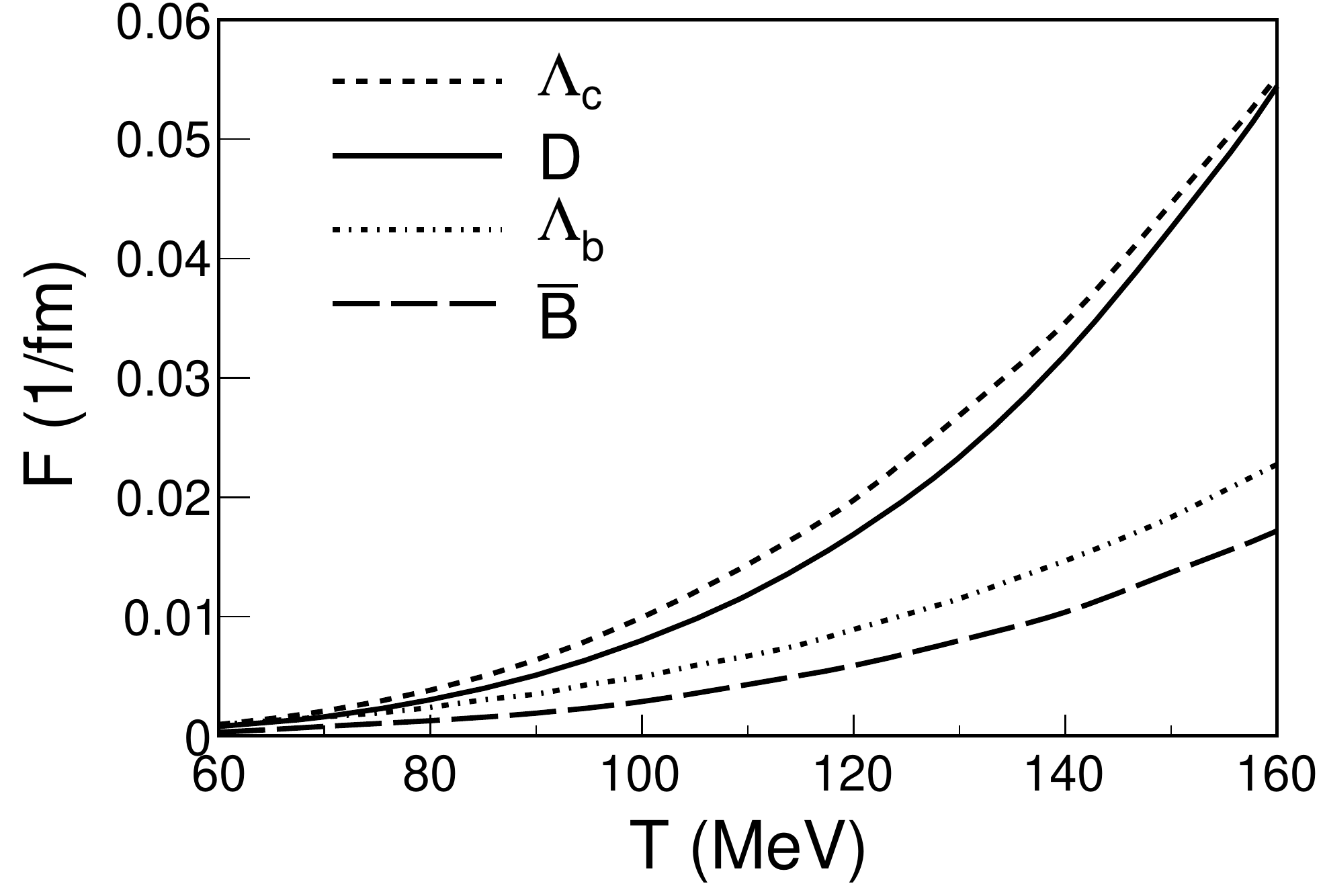}
\includegraphics[scale=0.4]{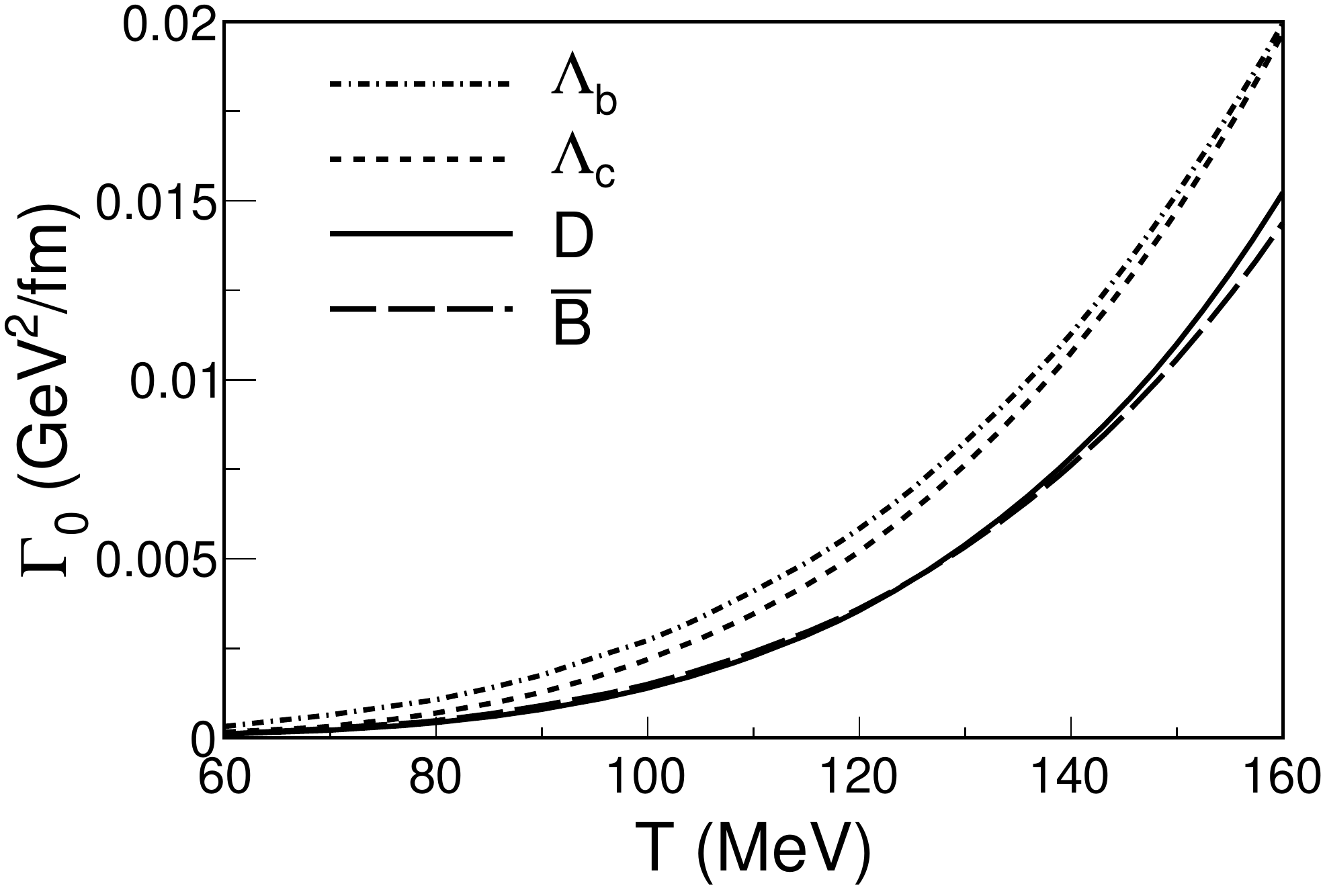}
\caption{\label{fig:allstates2}  
As in Fig.~\ref{fig:allstates} but in a gas of thermalized $\pi,K,\bar{K},\eta$ mesons.}
\end{center} 
\end{figure*}

 Once checked the consistency of our results with the nonrelativistic estimates, in Fig.~\ref{fig:allstates2} we present for completeness the same coefficients when all light mesons $(\pi,K,\bar{K},\eta)$ are included in the bath.
We also provide numerical tables of these results for their practical numerical application as supplementary files.

\subsection{Spatial diffusion coefficient}

The reduction of the Boltzmann or Boltzmann-Uehling-Uhlenbeck (BUU) equation to a Fokker-Planck equation is very convenient for heavy systems.
The main assumption is that the mass of the heavy particle propagating in the thermal bath is much bigger than the mass
of the surrounding particles and the temperature of the heat bath. In the Fokker-Planck approach, one does not need to solve the kinetic equation for the distribution function $f_p$. Instead, the explicit 
expressions of the transport coefficients in Eqs.~(\ref{eq:Fcoeff},\ref{eq:G0coeff},\ref{eq:G1coeff}) are obtained for
the first and second moments of the collision integral, without referring to $f_p$. Moreover, the
transport coefficients of the heavy particle are described in the momentum space rather than coordinate space.

The spatial diffusion coefficient $D_x$ contains information on how much the particles are disseminated in space. The mean
quadratic displacement ${\bf x}=(x,y,z)$ of the heavy particle as a function of time is approximately~\cite{risken1984fokker}
\be \langle ({\bf x}(t)-{\bf x}(t=0))^2 \rangle = 6 D_x t \ \ , \ee
where $D_x$ can be understood as the ``speed'' of the particle in a thermal medium. Within the Fokker-Planck approach, $D_x$ can be obtained from the diffusion coefficient in momentum space in the static limit ($p\rightarrow 0$),
\be D_x = \frac{\Gamma}{m_H^2 F^2} = \frac{T^2}{\Gamma} \ , \ee
where we have used Eq.~(\ref{eq:einstein}). The nonrelativistic estimate in Eq.~(\ref{eq:estimateG}) suggests that $D_x$ does not explicitly depend on the
heavy mass (but an indirect dependence is not precluded, for example through subleading effects in the cross section). 

In this work we compute the spatial diffusion coefficient (normalized by the de Broglie wavelength $(2\pi T)^{-1}$) for $\Lambda_c$ and $\Lambda_b$ baryons by means of the BUU equation and compare it to the  coefficient calculated within the Fokker-Planck framework. For the BUU calculation, we adapt the
calculation in Ref.~\cite{Torres-Rincon:2012sda}, where one of us computed the strangeness diffusion coefficient. The
charm and bottom diffusion coefficients are analogously performed following the same technique. In Appendix~\ref{appendix} we provide
some details on how the BUU equation is solved following the Chapman-Enskog expansion of the distribution function.

\begin{figure*}[htp]
\begin{center}
\includegraphics[scale=0.4]{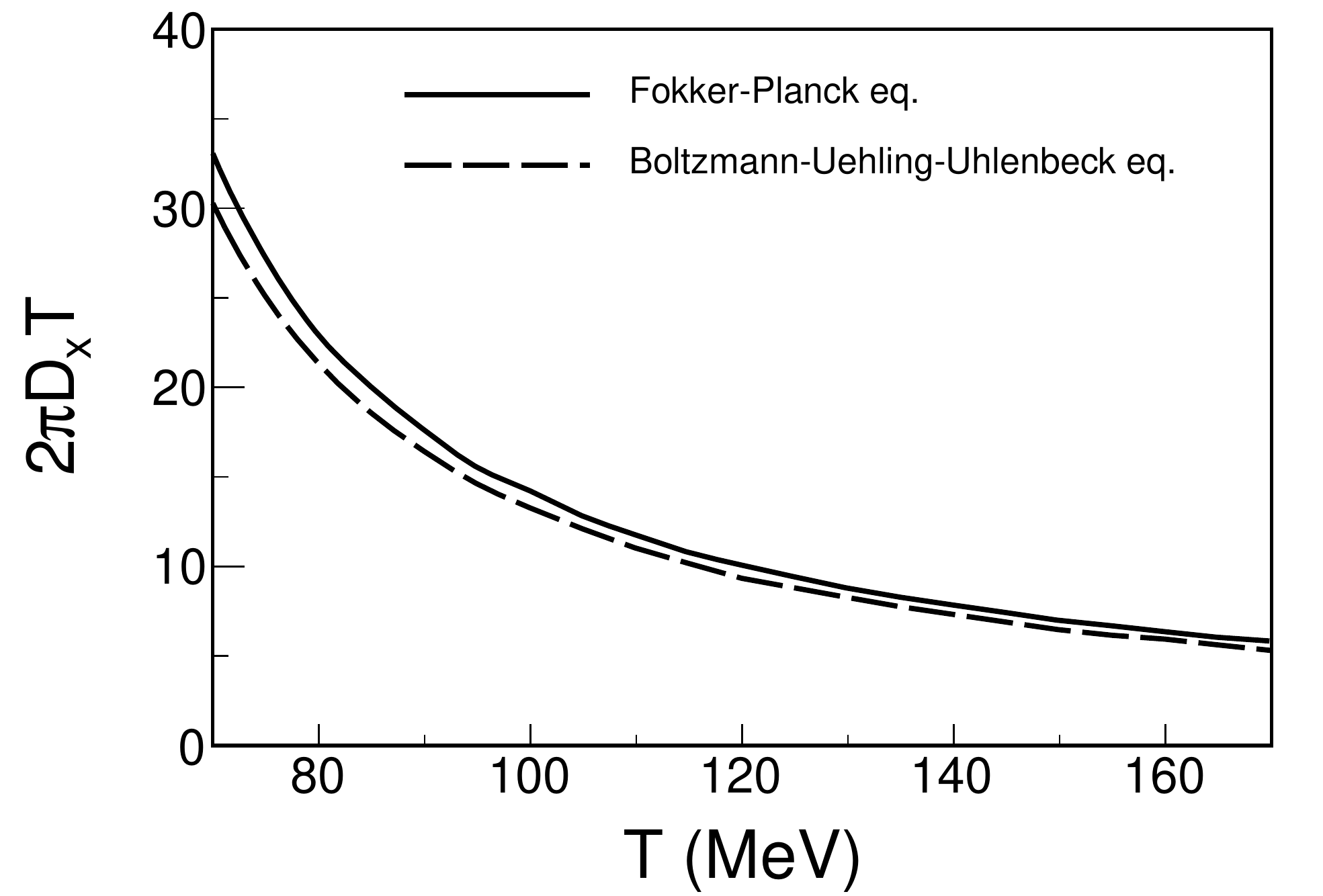}
\includegraphics[scale=0.4]{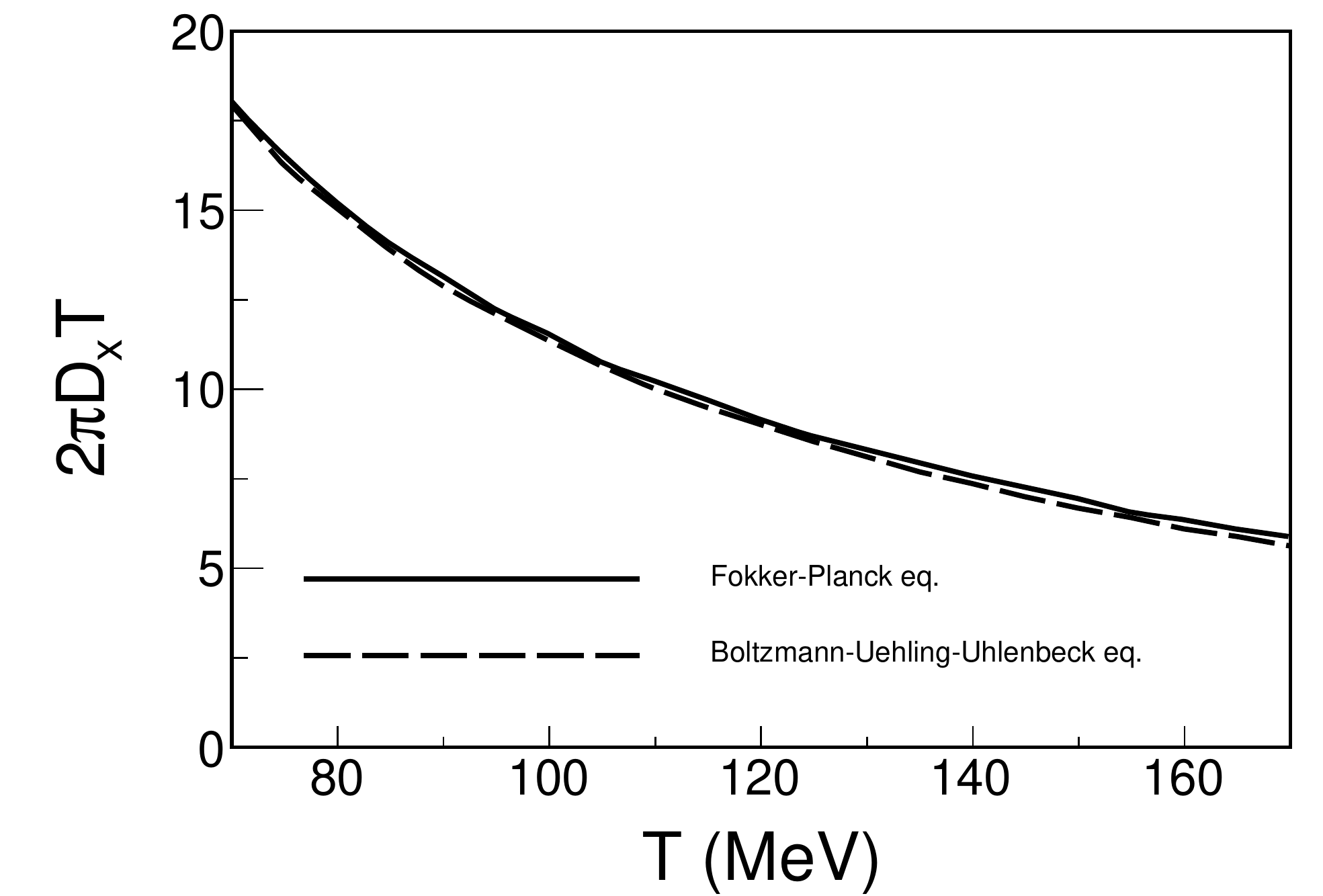}
\caption{\label{fig:diffusion}  $2\pi T D_x$ coefficient for $\Lambda_c$ (left panel) and $\Lambda_b$  (right panel) as a function of the temperature. }
\end{center} 
\end{figure*}

In Fig.~\ref{fig:diffusion} we  show our results for the spatial diffusion coefficient for both $\Lambda_c$ (left panel) and
$\Lambda_b$ (right panel). There is an excellent agreement between the Fokker-Planck and the BUU solutions in
spite of a completely different methodology. Thus, our results point to the validity of the Fokker-Planck approach
in order to extract the transport coefficients for the $\Lambda_c$ and $\Lambda_b$ baryons. We observe that the Fokker-Planck reduction works better for
$\Lambda_b$ than for $\Lambda_c$. This is indeed expected as the validity of the Fokker-Planck approach improves
when the mass difference between the heavy and the light particles increases. 

It should also be mentioned that in the BUU calculation one inverts the collision operator by performing a polynomial
expansion of the solution, which is truncated at some finite order (see Appendix~\ref{appendix} for details). This expansion converges exceptionally fast as it
can be seen in Table~\ref{tab:polyexp} for the diffusion coefficient at $T=140~{\rm MeV}$. In our results of Fig.~\ref{fig:diffusion} we have
considered the polynomial expansion up to third order.

\begin{table}
\begin{center}
\begin{tabular}{|c|c|c|c|}
\hline
$T=140$ MeV & \multicolumn{3}{|c|}{$2\pi TD_x$} \\ \cline{2-4}
\hline
Baryon & 1st order & 2nd order & 3rd order \\
\hline
\hline
$\Lambda_c$  & 7.2324 &        7.2375 &        7.2379   \\
$\Lambda_b$  & 7.3316 &       7.3326 &        7.3326   \\
\hline
\hline
\end{tabular}
\caption{\label{tab:polyexp} Convergence in the polynomial expansion of the 
solution of the BUU equation for the diffusion coefficient of  the $\Lambda_c$ and $\Lambda_b$ baryons at $T=140$ {\rm MeV}.}
\end{center}
\end{table}

The comparison between BUU and Fokker-Planck approaches has also been explored at the
level of the solution of the transport equation by one of us  for the heavy-quark dynamics in Ref.~\cite{Das:2013kea}. 
In that paper, it was found that the Fokker-Planck reduction (or the equivalent Langevin approach) is legitimate either
if the cross section is forward peaked (this is translated into a thermal Debye mass comparable with the
temperature), or if the heavy quark mass to temperature ratio is large ($m_H/T \ge 8-10$). 
For bottom quarks, the Langevin approach always gives very close results to those obtained by the Boltzmann transport
equation. For charm quarks, the Langevin dynamics brings considerable deviations depending on the value of the thermal Debye
mass, $m_D$. If $m_D$ is reduced, then the interaction is more forward peaked, thus enlarging the applicability
of the Fokker-Planck reduction. However, if $m_D$ is larger than $T$, then the Langevin approach provides very
different results from the Boltzmann equation. In our present approach the hadronic cross sections are not forward
peaked because all have been projected into the (isotropic) $s$-wave. However, we consider temperatures which are
much lower than those of the initial stages of the collisions. Therefore, the condition $m_H/T \gg 1$ is always
satisfied for heavy baryons. Then, we would expect that both formalisms provide similar results in the hadronic
evolution. 

Recent lattice-QCD calculations have reported results on the spatial diffusion coefficient at finite
temperature. As these calculations are performed for much higher temperatures than the ones analyzed here 
(already in the deconfined medium), one cannot expect the lattice-QCD outcome to be similar to our results, but
comparable with them at our top temperatures.
In Ref.~\cite{Banerjee:2011ra} the coefficient $2\pi T D_x$ is calculated for $T_c < T < 2T_c$ for heavy quarks
interacting with thermal gluons in the plasma, but not light quarks (quenched approximation). At $T=1.5 T_c$  a 
value of $2\pi TD_x =4.57 \pm 0.27 ^{+2.61}_{-1.56}$ is obtained. The authors of Ref.~\cite{Ding:2012sp} 
quote a smaller value of $2\pi TD_x (T = 1.46T_c)=1.8\pm0.7^{+1.3}_{-0.5}$ for charm without dynamical quarks.
Results in the continuum limit have recently appeared above the transition temperature~\cite{Francis:2015daa}. In
that work, the mass of the heavy probe is taken to be asymptotically high ($M_H \gg \pi T$) and the spatial diffusion 
coefficient is computed again in a pure SU(3) gauge theory. The value obtained in the continuum limit 
reads $2\pi TD_x \simeq 3.7-6.9$ around $T \sim 1.5T_c$ (with $T_c \simeq 317$ MeV). These values from the lattice-QCD
calculations seem to be compatible with the tendency of our results at the highest temperatures (but always below $T_c$).

\section{Conclusions~\label{sec:conclusions}}

  In this work we have reported our findings of the transport coefficients of $\Lambda_c$ and $\Lambda_b$ baryons
immersed in a hot hadronic medium. Although transport properties of heavy mesons $(D^{(*)},{\bar B}^{(*)})$ have been widely 
considered in the literature~\cite{Laine:2011is,He:2011yi,Ghosh:2011bw,Abreu:2011ic,Das:2011vba,Tolos:2013kva,Torres-Rincon:2014ffa}, 
the energy loss and momentum diffusion of heavy baryons
have received little attention so far (see Ref.~\cite{Ghosh:2014oia} for a first study on $\Lambda_c$ transport coefficients).

  We have used a low-energy effective Lagrangian based on chiral and heavy-quark spin symmetries to describe the 
interaction of these states with light mesons ($\pi, K, {\bar K}, \eta$). In order to obtain the physical scattering
amplitudes we have unitarized the interaction kernels
using a Bethe-Salpeter equation in the ``on-shell'' approximation. The resulting scattering amplitudes respect
unitarity bounds and, thus, provide sensible values for the transport cross sections. As a byproduct, we have
generated several resonant states (poles of the scattering amplitudes) that dominate the hadron scattering at
low energies.

  These cross sections have been implemented to compute the drag force and diffusion coefficients for $\Lambda_c$ and $\Lambda_b$ as
a function of the temperature and heavy-baryon momentum at vanishing baryochemical potential. To perform the calculation we have used the
Fokker-Planck approximation of the transport equation, having immediate access to these coefficients
as a function of the heavy-baryon momentum. We have presented intermediate checks of our results: 1) The
Einstein relation is satisfied for all temperatures. 2) The nonrelativistic limits and scaling relations between
coefficients of different species (including $D$ and ${\bar B}$ mesons) are satisfied in a good degree of accuracy. 3) 
The spatial diffusion coefficient $D_x$ has been checked against the one calculated
using the Boltzmann-Uehling-Uhlenbeck transport equation. Our results for the diffusion
coefficients follow the same trend as those for heavy mesons, and they are fully compatible with lattice-QCD results
of deconfined heavy matter at the edge of the phase transition temperature.

   With these results in hand, we can consider the full evolution of heavy quarks+heavy baryons in a hot medium,
simulating the heavy-flavor diffusion in high-energy heavy-ion collisions. This requires to solve a transport equation 
for the evolution of the particle distribution as a function of time (or in the Langevin approach, the dynamics of 
individual particles to know their trajectories in time). A detailed calculation of the $\Lambda_c$ and
$\Lambda_b$ observables will be reported in a subsequent article~\cite{Das:2016llg}. We will provide results for the 
standard observables characterizing the medium modification of the hadrons like $R_{AA}$, $v_2$
and baryon to meson ratios ($\Lambda_c/D$ and $\Lambda_b/\bar{B}$) for high-energy collisions at LHC and RHIC.
While the heavy quark evolution has been considered previously in a variety of schemes, the baryon evolution after the hadronization into
$\Lambda_c$ and $\Lambda_b$ will be implemented using the effective interaction that we have
detailed here.

Experimental measurements for the $\Lambda_c$ observables are foreseen at the Run 3 by the ALICE
Collaboration~\cite{Abelevetal:2014dna,Andronic:2015wma}, with little attention to the $\Lambda_b$. With the present paper we are able to study the dynamics of $\Lambda_c$ and $\Lambda_b$ states in a hot medium using realistic interactions that can be checked against experiment.  Moreover, the transport properties of  these states can be calculated in a very confident way, as we have found that there exists full consistency between Boltzmann and Fokker-Planck approaches. Finally, the fact that heavy baryons, such as  $\Lambda_c$ and $\Lambda_b$, carry an extra conserved quantity, the baryon number, opens up the possibility to study the interplay of the different transport channels (in the Onsager's sense of a mixed heavy flavor--baryon diffusion coefficients ~\cite{onsager}). For all these
reasons, we encourage the experimental collaborations to pursue the study of heavy baryons in heavy-ion collisions in
the future, in spite of the well-known technical difficulties for their reconstruction.

\section{Acknowledgements}

This work has been financed by Grants No. FPA2010-16963 and No. FPA2013-43425-P (Spain),
EU Integrated Infrastructure Initiative Hadron Physics Project under Grant
Agreement n. 227431 and Grant No. FP7-PEOPLE-2011-CIG under Contract No. PCIG09-GA-2011-291679.

J.M.T.-R. acknowledges financial support from programme TOGETHER from R\'egion Pays de la Loire.
L.T. acknowledges support from the Ram\'on y Cajal research programme from Ministerio de Econom\'{\i}a y Competitividad.
S.K.D. acknowledges the support by the ERC StG under the QGPDyn Grant no. 259684.

\appendix

\section{Computation of $D_x$ from the Boltzmann-Uehling-Uhlenbeck equation\label{app:BUU}}
\label{appendix}
  In this Appendix we sketch the calculation of the spatial diffusion coefficient $D_x$ using the BUU Eq.~(\ref{eq:BUU}) instead of the Fokker-Planck approach.
Being a more general equation, the solution of the BUU equation provides a consistency check for our results as well
as an applicability test of the Fokker-Planck reduction. Not being a central part of the present work, we only give a general overview of the procedure. For more details we refer to Ref.~\cite{Torres-Rincon:2012sda}, where the method
was applied to compute the strangeness diffusion. 

  The spatial diffusion coefficient can be defined in terms of the Fick diffusion law as
\be \label{eq:fick} {\bf j} (t,{\bf x})= - D_x \ \nabla j^0 (t,{\bf x})  \ , \ee    
where $j^0$ and ${\bf j}$ are respectively the concentration and density current of heavy particles. In terms of a four-vector 
we have $j^\mu = (j^0,{\bf j})$. This phenomenological law is valid as long as higher order
gradients can be neglected. For this reason we always consider the system close to the equilibrium state, where
gradients are small.

 The connection with microphysics is given by the expression of the current in terms of the distribution
function $f_p(t,{\bf x})$,
\be \label{eq:4current} j^\mu (t,{\bf x}) = g \int \frac{d {\bf p}}{(2\pi)^3 2E_p}  f_p (t,{\bf x}) \ (2p^\mu) \ , \ee
where $g$ is the degeneracy of heavy baryons and $E_p=p^0=\sqrt{p^2+m_H^2}$. Close to equilibrium, the distribution
function obeys the BUU equation~(\ref{eq:BUU})
\be \label{eq:BUU2} \frac{df_p}{dt} = g_l \int_{k,q} \ d\Gamma \
 [ \ f_{p-k} n_{q+k}  (1 - f_p) (1 + n_q )  - f_p n_q (1 - f_{p-k}) (1 + n_{q+k}) \ ] \ , \ee
where $d\Gamma = d\Gamma_{p,q \rightarrow p-k,q+k}$, $g_l$ is the degeneracy factor of the light particle~\footnote{The generalization to many species is straighforward as one only needs to consider a sum in $l$. We remind
that in this work $l\in \pi,K,{\bar K},\eta$}. The light system is already taken in equilibrium, with
the distribution function $n_p$ being the Bose-Einstein function. We also recall that the momentum label specifies
the particle involved, being $q$ and $q+k$ related to the light particle 
and momenta $p$ and $p-k$ related to the heavy particle.

To solve the BUU equation we use the Chapman-Enskog expansion~\cite{chapman1952mathematical}, which is one of the standard techniques when the system is not far from equilibrium.
This method assumes that the distribution function can be expanded in powers of the Knudsen number, $Kn=\lambda_{\rm mfp}/L_{\rm macro}$, with $\lambda_{\rm mfp}$ the mean-free path of
the heavy particle and $L_{\rm macro}$ a typical ``inhomogeneity scale'' of the system~\cite{chapman1952mathematical,de1980relativistic,Torres-Rincon:2012sda}. Equivalently, the 
expansion can be regarded as an expansion in powers of gradients of the hydrodynamic functions (temperature, chemical potential, velocity field...)~\cite{chapman1952mathematical,de1980relativistic}.
In the present case we  only consider the gradient in chemical potential $\nabla \mu$. In addition, the gradient is always small enough to keep a linear approximation.

The expansion in the distribution function of the heavy particle reads
\be \label{eq:CE} f_p = n_p + f^{(1)} (p) + ... \ , \ee
where $n_p$ is the Fermi-Dirac distribution function, and the first correction, $f^{(1)} (p)$, is ${\cal O} (Kn^1)$ (or first order in gradients ${\cal O} (\nabla)$). 

We now introduce the expansion (\ref{eq:CE}) into the BUU equation. In the left-hand side (LHS), we assume that any
spacetime dependence of $f_p (t,{\bf x})$ comes only through inhomogeneities of the (heavy) chemical potential,
\be \frac{df_p}{dt} = \frac{d}{dt} n_p ( \mu({\bf x})) + {\cal O} (\nabla^2)=  n_p (1 - n_p) \beta \frac{{\bf p}}{E_p} \cdot  \nabla \mu + {\cal O}(\nabla^2) \ , \ee
where $\beta=1/T$.

  In the right-hand-side (RHS) we introduce the same expansion (\ref{eq:CE}) and linearize the collision operator.
Further, we use the ansatz $f^{(1)} (p)=-n_p (1 - n_p) {\bf p} \cdot \nabla \mu \beta^3 H(p)$ to make the gradient 
in chemical potential explicit. We thus find an equation for $H(p)$,
\be \label{eq:linearBUU}  n_p (1 - n_p) {\bf p}  = - g_l E_p \beta^2 \int_{k,q} \ d\Gamma \
n_{p-k} n_{q+k} (1 + n_{q}) (1- n_{p}) [ {\bf p} H(p) - ({\bf p-k}) H (p-k)] \ . \ee
The integral equation is linear in $H(p)$, thus being an important simplification of the nonlinear collision operator in
Eq.~(\ref{eq:BUU2}). 
In addition, one has the advantage that the LHS does not depend on $H(p)$, easing the inversion of the collision
integral. The procedure is simplified by expanding this function in terms of a convenient orthogonal polynomial basis $\{ P_i (p) \}$,
\be \label{eq:expand} H(p) = \sum_{i} \mathfrak{h}_i P_i (p) \ , \ee
where the unkown coefficients $\mathfrak{h}_i$ are independent of $p$ (but temperature and density dependent). The polynomial
basis can be straighforwardly constructed following a Gram-Schmidt method once an integration measure is given. The
integration measure $d\mu(p)$ (not to be confused with the chemical potential) is chosen as
\be \label{eq:measure} d\mu(p) \equiv \frac{dp}{E_p} n_p (1 - n_p) \frac{p^4}{m_H^4} \  , \ee
with momentum $p \in [0,+\infty)$. The integration measure induces a scalar product, and the concept of
orthogonality needed for the construction of the polynomial basis. For more details, we refer
again to~\cite{Torres-Rincon:2012sda} where all these ingredients are defined to account for the Hilbert-space structure in which the solution function
$H(p)$ lives.

The transport equation~(\ref{eq:linearBUU}) is projected multiplying by $P_n (p) dp p^2 p^i /(E_p m_H^4)$ and integrating over momentum in both sides
\be \int d\mu(p)P_n(p) = - \frac{g_l}{m_H^4 T^2} \sum_i \mathfrak{h}_i  \int_{p,k,q} \ dp p^2 \ d\Gamma \
n_{p-k} n_{q+k} (1+ n_{q}) (1- n_{p})  {\bf p} P_n(p) [ {\bf p} P_i(p) - ({\bf p-k}) P_i (p-k)] . \ee
Although this equation might seem still complicated, one should recall that the only unknown coefficients are  $\mathfrak{h_i}$. Defining
\be {\cal C}_{ni} \equiv - \frac{g_l}{m_H^4 T^2}  \int_{p,k,q} \ dp p^2 \ d\Gamma \ 
n_{p-k} n_{q+k} (1+ n_{q}) (1 - n_{p}) {\bf p}  P_n (p) [ {\bf p} P_i(p) - ({\bf p-k}) P_i (p-k)] \ , \ee
the solution for $H(p)$ is obtained order by order in the polynomial expansion as
\be \label{eq:solh} \mathfrak{h}_i =  \sum_n [{\cal C}^{-1}]_{in} \ \left[ \int d\mu(p) P_n(p) \right] \ , \ee

In the RHS of the Fick's law (\ref{eq:fick}), the expansion (\ref{eq:CE}) can be truncated at zeroth order, as it is
explicitly linear in gradients,
\be \label{eq:grad} \nabla j^0 (t,{\bf x}) = g \int \frac{d {\bf p}}{(2\pi)^3} \nabla f_p (t,{\bf x}) =  g \int \frac{d {\bf p}}{(2\pi)^3}  n_p (1 - n_p) \beta \nabla \mu ({\bf x})+ {\cal O} (\nabla^2) \ , \ee  
where we have used Eq.~(\ref{eq:4current}). However, in the LHS one should consider the term with $f^{(1)}$, because the zeroth-order term vanishes due to the symmetry of the integral. The Fick equation is then rewritten in terms of $H(p)$ as 
\be \label{eq:coeff} \int \frac{d {\bf p}}{(2\pi)^3} \frac{{\bf p}}{E_p} n_p (1- n_p) {\bf p} \cdot \nabla \mu \beta^2 H(p)=
D_x {\nabla \mu} \int \frac{d {\bf p}}{(2\pi)^3} n_p (1 - n_p) \ . \ee

Expressing the integrations as scalar products with the integration measure of Eq.~(\ref{eq:measure}), we can obtain 
the diffusion coefficient
\be D_x =  \frac{1}{3T^2} \frac{\int d\mu(p) \ H(p)}{ \int d\mu(p) \ E_p/p^2 } =
\frac{1}{3T^2} \frac{1}{ \int d\mu(p) \ E_p/p^2 } \sum_i \mathfrak{h}_i \int d\mu(p) \ P_i(p)  \ , \ee
where we have introduced (\ref{eq:expand}). 
Finally, we substitute the solution of the transport equation to obtain
\be 2\pi T D_x =  \frac{2\pi}{3T}  \frac{1}{ \left[ \int d\mu(p) \ E_p/p^2 \right] } \  \sum_{i,n} 
 \left[ \int d\mu(p) \ P_i(p) \right]  [{\cal C}^{-1}]_{in}  \left[ \int d\mu(p) P_n(p) \right]  \ .\ee

By choosing an adequate polynomial basis, we can solve the previous equation with a few terms of the polynomial expansion.
We refer to~\cite{Torres-Rincon:2012sda} for additional information, while a more detailed calculation will be reported elsewhere.

\end{document}